\documentclass[preprint,amsmath,amssymb,aps,sort&compress,merge,prd,nofootinbib]{revtex4}

\usepackage{amsmath}   % for split
\usepackage{graphicx}  % needed for figures
\usepackage{float}
\usepackage{dcolumn}   % needed for some tables
\usepackage{bm}        % for math
\usepackage{slashed}   % for Dirac Slash
\usepackage{simplewick} % for contraction
\usepackage{verbatim}  % for multi-line comment
\usepackage{color} % for textcolor
\usepackage{appendix} % for appendix
\usepackage{subfigure}

\graphicspath{{figure/}}

\newcounter{JW}

\usepackage[bookmarks=true,colorlinks=true,linkcolor=blue,unicode=true]{hyperref}

\begin{document}

%\widetext

\title{Constraints on anomalous quartic gauge couplings via $Z\gamma jj$ production at the LHC}

\vskip 0.25cm

\date{\today}

\author{Ji-Chong Yang}
\email{yangjichong@lnnu.edu.cn}
\author{Yu-Chen Guo}
\email{ycguo@lnnu.edu.cn}
\author{Chong-Xing Yue}
\email{cxyue@lnnu.edu.cn}
\author{Qing Fu}
\affiliation{Department of Physics, Liaoning Normal University, Dalian 116029, China}

\begin{abstract}
In this paper, we investigate the contributions of anomalous quartic gauge couplings~(aQGCs) to $Z\gamma jj$ production at the Large Hadron Collider~(LHC) in the context of Standard Model effective theory~(SMEFT).
When energy scale is large, the validity of SMEFT becomes an important issue.
To ensure the validity, the unitarity bound is applied in a model independent approach, which is found to have significant suppressive effects on the signals of $O_{M_i}$ operators.
The kinematic and polarization features of the aQGC signals are also studied.
The polarization effect is useful to highlight the signals of $O_{T_i}$ operators.
The sensitivity estimates on dimension-8 operators with unitarity bounds at $\sqrt{s}=14$ TeV are obtained.
\end{abstract}

\maketitle

\section{\label{level1}Introduction}

The extension of gauge interactions has been attracting widespread attention in the search of new physics~(NP) beyond the Standard Model~(SM).
From the perspective of the SM effective field theory~(SMEFT)~\cite{weinberg,*SMEFTReview1,*SMEFTReview2,*SMEFTReview3}, up to dimension-8, modifications to the gauge interactions can come from high-dimensional operators contributing to anomalous trilinear gauge couplings~(aTGCs) and anomalous quartic gauge couplings~(aQGCs).
aTGCs can only be generated from loop diagrams while aQGCs can originate from tree diagrams~\cite{looportree}.
Besides, aQGCs can lead to richer helicity combinations than aTGCs~\cite{ssww}, thus it is possible that the interference between aQGCs and the SM is not suppressed as those of aTGCs~\cite{atgcsuppresed,*helicitySuppress}.
Note that the interference between dimension-8 aQGCs and the SM is at the same order as the square terms of the dimension-6 operators in the EFT expansion.
Therefore aQGCs are phenomenologically important.
Since aQGCs and aTGCs from dimension-6 operators are not independent of each other~\cite{vbscan}, it is necessary to start with the dimension-8 operators if one focuses on only aQGCs.
There are many NP models that contain effective dimension-8 operators contributing to aQGCs, such as composite Higgs~\cite{composite1,*composite2}, warped extra dimensions~\cite{extradim}, 2 Higgs doublet models~\cite{2hdm1,*2hdm2}, $U(1)_{L_{\mu}-L_{\tau}}$~\cite{zprime1,*zprime2}, as well as axion-like particles~\cite{alp1,*alp2}.
In some specific cases, dimension-6 operators are absent but the dimension-8 operators will show up, for example the Born-Infeld theory~\cite{bi1,*bi2} and the neutral triple gauge couplings~\cite{ntgc1,*ntgc2,*ntgc3}.
Moreover, recent studies have shown that the dimension-8 operators play a very important role from the convex geometry perspective to the SMEFT space~\cite{convexgeometry}.
As a result, dimension-8 operators contributing to aQGCs have been intensively studied in the last few years~\cite{jrr1,jrr2,*Guo:2019agy,*5gaugeinteraction}.

The vector boson scattering processes~(VBS) at the Large Hadron Collider~(LHC) are very suitable for studying aQGCs~\cite{vbs1,*vbs3,*vbs4}.
The same-sign WW production is the first reported VBS process at the LHC and is also the first evidence of QGCs~\cite{sswwexp1}.
Since then, a number of VBS processes have been investigated, including $Z\gamma jj$~\cite{zaexp1,zaexp2,zaexp3}, $W\gamma jj$~\cite{waexp1}, $ZZjj$~\cite{zzexp1,*zzexp2}, $WZjj$~\cite{wzexp1,*wzexp2} and $WWjj$~\cite{wwexp1,wwexp2}.
Among those VBS processes, the $Z\gamma jj$ production can be affected by $WWZ\gamma$, $ZZZ\gamma$, $Z\gamma\gamma\gamma$ and $ZZ\gamma\gamma$ couplings.
The next-to-leading order~(NLO) QCD correction to the $pp\to Z\gamma jj$ in the SM has been computed in Refs.~\cite{zanlo,vbfcut}, and the K-factor is found to be close to one ($K\approx 1.02$~\cite{vbfcut}).
In this paper, we focus on the phenomenology of this channel with aQGCs at $14$ TeV LHC.

As an effective theory, SMEFT only applies within a specific energy scale.
Unitarity is often used to determine whether SMEFT is valid.
Unitarization methods such as K-matrix~\cite{jrr2}, form factors~\cite{wwexp1,vbs1,vbfcut} and dispersion relations~\cite{dispersion1,*dispersion2} can be introduced to avoid the violation of unitarity.
However, the constraints on the effective operators dependent on the unitarization method used~\cite{unitarizationeffects}.
In order to compare with each other, many experiments extracting the coefficients of aQGCs did not use unitarization method.
In this paper, we do not use unitarization methods but use a matching procedure proposed in Refs.~\cite{matchingidea2,*matchingidea3,wastudy} which can be used in experiments.
Based on the study of the kinematic features, and the angular distributions due to the polarization features, we optimize the event selection strategy specifically for aQGCs in the process $pp\to jj \ell^+\ell^-\gamma$.
The constraints on the coefficients are presented by using Monte-Carlo~(MC) simulation.

The remainder of the paper is organized as follows. In Sec.~\ref{level2}, we establish the aQGCs in dimension-8 operators relevant to the process $pp\to jj\ell^+\ell^- \gamma$. In Sec.~\ref{level3} we analyse the partial-wave unitarity bounds in the $VV \to Z\gamma$ subprocesses and introduce the matching procedure. In Sec.~\ref{level4}, the features of the aQGC signals are presented, and the event selection strategy is discussed. The constraints on the coefficients with unitarity bounds applied are shown in Sec.~\ref{level5}. Sec.~\ref{level6} is a summary.

\section{\label{level2} Operator basis}

The dimension-8 operators contributing to aQGCs relevant to the process $pp\to jj \ell ^+ \ell^- \gamma$ can be written as~\cite{aqgcold,*aqgcnew}
\begin{equation}
\begin{split}
&\mathcal{L}_{\rm aQGC}=\sum _{i} \frac{f_{M_i}}{\Lambda^4}O_{M,i}+\sum _{j} \frac{f_{T_j}}{\Lambda^4}O_{T,j},
\end{split}
\label{eq.2.1}
\end{equation}
with
\begin{equation}
\begin{split}
&O_{M,0}={\rm Tr\left[\widehat{W}_{\mu\nu}\widehat{W}^{\mu\nu}\right]}\times \left[\left(D^{\beta}\Phi \right) ^{\dagger} D^{\beta}\Phi\right],\\
&O_{M,1}={\rm Tr\left[\widehat{W}_{\mu\nu}\widehat{W}^{\nu\beta}\right]}\times \left[\left(D^{\beta}\Phi \right) ^{\dagger} D^{\mu}\Phi\right],\\
&O_{M,2}=\left[B_{\mu\nu}B^{\mu\nu}\right]\times \left[\left(D^{\beta}\Phi \right) ^{\dagger} D^{\beta}\Phi\right],\\
&O_{M,3}=\left[B_{\mu\nu}B^{\nu\beta}\right]\times \left[\left(D^{\beta}\Phi \right) ^{\dagger} D^{\mu}\Phi\right],\\
&O_{M,4}=\left[\left(D_{\mu}\Phi \right)^{\dagger}\widehat{W}_{\beta\nu} D^{\mu}\Phi\right]\times B^{\beta\nu},\\
&O_{M,5}=\left[\left(D_{\mu}\Phi \right)^{\dagger}\widehat{W}_{\beta\nu} D_{\nu}\Phi\right]\times B^{\beta\mu} + h.c.,\\
&O_{M,7}=\left(D_{\mu}\Phi \right)^{\dagger}\widehat{W}_{\beta\nu}\widehat{W}_{\beta\mu} D_{\nu}\Phi,\\
\end{split}
\label{eq.2.2}
\end{equation}
\begin{equation}
\begin{split}
&O_{T,0}={\rm Tr}\left[\widehat{W}_{\mu\nu}\widehat{W}^{\mu\nu}\right]\times {\rm Tr}\left[\widehat{W}_{\alpha\beta}\widehat{W}^{\alpha\beta}\right],\\
&O_{T,1}={\rm Tr}\left[\widehat{W}_{\alpha\nu}\widehat{W}^{\mu\beta}\right]\times {\rm Tr}\left[\widehat{W}_{\mu\beta}\widehat{W}^{\alpha\nu}\right],\\
&O_{T,2}={\rm Tr}\left[\widehat{W}_{\alpha\mu}\widehat{W}^{\mu\beta}\right]\times {\rm Tr}\left[\widehat{W}_{\beta\nu}\widehat{W}^{\nu\alpha}\right],\\
&O_{T,5}={\rm Tr}\left[\widehat{W}_{\mu\nu}\widehat{W}^{\mu\nu}\right]\times B_{\alpha\beta}B^{\alpha\beta},\\
&O_{T,6}={\rm Tr}\left[\widehat{W}_{\alpha\nu}\widehat{W}^{\mu\beta}\right]\times B_{\mu\beta}B^{\alpha\nu},\\
&O_{T,7}={\rm Tr}\left[\widehat{W}_{\alpha\mu}\widehat{W}^{\mu\beta}\right]\times B_{\beta\nu}B^{\nu\alpha},\\
&O_{T,8}=B_{\mu\nu}B^{\mu\nu}\times B_{\alpha\beta}B^{\alpha\beta},\\
&O_{T,9}=B_{\alpha\mu}B^{\mu\beta}\times B_{\beta\nu}B^{\nu\alpha},\\
\end{split}
\label{eq.2.3}
\end{equation}
where $\widehat{W}\equiv \sum _i W^i \sigma ^i / 2$. The constraints on the coefficients obtained from the experiments are listed in Table.~\ref{tab.1}.
\begin{table}
\begin{center}
\begin{tabular}{c|c ||c|c}
\hline
coefficient & constraint & coefficient & constraint\\
\hline
$f_{M_0}/\Lambda ^4$ & $[-0.69, 0.70]$~\cite{coefficient1} & $f_{T_0}/\Lambda ^4$ & $[-0.12, 0.11]$~\cite{coefficient1} \\
$f_{M_1}/\Lambda ^4$ & $[-2.0, 2.1]$~\cite{coefficient1} & $f_{T_1}/\Lambda ^4$ & $[-0.12, 0.13]$~\cite{coefficient1} \\
$f_{M_2}/\Lambda ^4$ & $[-8.2, 8.0]$~\cite{zaexp3} & $f_{T_2}/\Lambda ^4$ & $[-0.28, 0.28]$~\cite{coefficient1} \\
$f_{M_3}/\Lambda ^4$ & $[-21, 21]$~\cite{zaexp3} & $f_{T_5}/\Lambda ^4$ & $[-0.7, 0.74]$~\cite{zaexp3} \\
$f_{M_4}/\Lambda ^4$ & $[-15, 16]$~\cite{zaexp3} & $f_{T_6}/\Lambda ^4$ & $[-1.6, 1.7]$~\cite{zaexp3} \\
$f_{M_5}/\Lambda ^4$ & $[-25, 24]$~\cite{zaexp3} & $f_{T_7}/\Lambda ^4$ & $[-2.6, 2.8]$~\cite{zaexp3} \\
$f_{M_7}/\Lambda ^4$ & $[-3.4, 3.4]$~\cite{coefficient1} & $f_{T_8}/\Lambda ^4$ & $[-0.47, 0.47]$~\cite{zaexp3} \\
& & $f_{T_9}/\Lambda ^4$ & $[-1.3, 1.3]$~\cite{zaexp3} \\
\hline
\end{tabular}
\end{center}
\caption{\label{tab.1}The constraints on the coefficients~(${\rm TeV^{-4}}$) obtained by experiments.}
\end{table}

The aQGCs relevant to the process $pp\to jj \ell ^+ \ell^- \gamma$ are $\gamma ZWW$, $\gamma \gamma ZZ$, $\gamma ZZZ$ and $\gamma \gamma \gamma Z$ couplings, the $\gamma Z WW$ couplings are given in Ref.~\cite{wastudy}, and the others are listed below,
%\begin{equation}
%\begin{split}
%&V_{AZWW,0}=F^{\mu\alpha}Z_{\mu\beta}(W^+_{\alpha}W^{-\beta}+W^-_{\alpha}W^{+\beta}),\\
%&V_{AZWW,2}=F^{\mu\nu}Z_{\mu\nu}W^+_{\alpha}W^{-\alpha},\\
%&V_{AZWW,4}=F^{\mu\alpha}Z^{\beta}(W^+_{\mu\beta}W^-_{\alpha}+W^-_{\mu\beta}W^+_{\alpha}),\\
%&V_{AZWW,6}=F^{\mu\alpha}Z_{\mu\beta}(W^+_{\nu\alpha}W^{-\nu\beta}+W^-_{\nu\alpha}W^{+\nu\beta}),\\
%\end{split}
%\quad
%\begin{split}
%&V_{AZWW,1}=F^{\mu\alpha}Z_{\alpha}(W^+_{\mu\beta}W^{-\beta}+W^-_{\mu\beta}W^{+\beta}),\\
%&V_{AZWW,3}=F^{\mu\alpha}Z^{\beta}(W^+_{\mu\alpha}W^-_{\beta}+W^-_{\mu\alpha}W^+_{\beta}),\\
%&V_{AZWW,5}=F^{\mu\nu}Z_{\mu\nu}W^{+\alpha\beta}W^-_{\alpha\beta},\\
%&V_{AZWW,7}=F^{\mu\nu}Z^{\alpha\beta}(W^+_{\mu\nu}W^-_{\alpha\beta}+W^-_{\mu\nu}W^+_{\alpha\beta}),
%\end{split}
%\label{eq.2.4}
%\end{equation}
\begin{equation}
\begin{split}
&V_{2A2Z,0}=F^{\mu\nu}F_{\mu\nu}Z^{\alpha}Z_{\alpha},\\
&V_{2A2Z,2}=F^{\mu\nu}F_{\mu\nu}Z^{\alpha\beta}Z_{\alpha\beta},\\
&V_{2A2Z,4}=F^{\mu\nu}F_{\alpha\beta}Z^{\mu\nu}Z_{\alpha\beta},\\
\end{split}
\quad
\begin{split}
&V_{2A2Z,1}=F^{\mu\nu}F_{\mu\alpha}Z^{\nu}Z_{\alpha},\\
&V_{2A2Z,3}=F^{\mu\nu}F_{\nu\alpha}Z^{\nu\alpha}Z_{\alpha\beta},\\
&V_{2A2Z,5}=F^{\mu\nu}F_{\alpha\beta}Z_{\mu\beta}Z^{\alpha\nu},\\
\end{split}
\label{eq.2.5}
\end{equation}
\begin{equation}
\begin{split}
&V_{A3Z,0}=F^{\mu\nu}Z_{\mu\nu}Z^{\alpha}Z_{\alpha},\\
&V_{A3Z,2}=F^{\mu\nu}Z_{\mu\nu}Z^{\alpha\beta}Z_{\alpha\beta},\\
\end{split}
\quad
\begin{split}
&V_{A3Z,1}=F^{\mu\nu}Z_{\mu\alpha}Z^{\nu}Z_{\alpha},\\
&V_{A3Z,3}=F^{\mu\alpha}Z_{\mu\beta}Z^{\nu\alpha}Z_{\nu\beta},\\
\end{split}
\label{eq.2.6}
\end{equation}
\begin{equation}
\begin{split}
&V_{3AZ,0}=F^{\mu\nu}F_{\mu\nu}F^{\alpha\beta}Z_{\alpha\beta},\\
\end{split}
\quad
\begin{split}
&V_{3AZ,1}=F^{\mu\nu}F_{\nu\alpha}F^{\alpha\beta}Z_{\beta\mu},\\
\end{split}
\label{eq.2.7}
\end{equation}
with $W^{\pm \mu\nu}\equiv \partial^{\mu} W^{\pm  \nu} -\partial^{\nu} W^{\pm  \mu}$ and $Z^{\mu\nu}\equiv \partial^{\mu} Z^{\nu} -\partial^{\nu} Z^{\mu}$. The coefficients are
%\begin{equation}
%\begin{split}
%&\alpha_{AZWW,0}=\frac{e^2v^2}{8\Lambda ^4}\left(\frac{c_W^2}{s_W^2}f_{M_5}-f_{M_5}-\frac{c_W}{s_W}f_{M_1}\right.\\
%&\left.+2\frac{c_W}{s_W}f_{M_3}+\frac{c_W}{2s_W}f_{M_7}\right),\\
%&\alpha_{AZWW,1}=\frac{e^2v^2}{8\Lambda ^4}\left(-\frac{1}{2c_Ws_W}f_{M_7}-\frac{1}{s_W^2}f_{M_5}\right),\\
%&\alpha_{AZWW,2}=\frac{e^2v^2}{8\Lambda ^4}\left(\frac{c_W^2}{s_W^2}f_{M_4}-f_{M_4}+2\frac{c_W}{s_W}f_{M_0}-4\frac{c_W}{s_W}f_{M_2}\right),\\
%&\alpha_{AZWW,3}=-\frac{e^2v^2}{8\Lambda ^4}\frac{1}{s_W^2}f_{M_4},\\
%&\alpha_{AZWW,4}=\frac{e^2v^2}{8\Lambda ^4}\left(\frac{1}{2c_Ws_W}f_{M_7}-\frac{1}{s_W^2}f_{M_5}\right),\\
%&\alpha_{AZWW,5}=\frac{2c_Ws_W}{\Lambda^4}\left(f_{T_0}-f_{T_5}\right),\\
%&\alpha_{AZWW,6}=\frac{c_Ws_W}{\Lambda^4}\left(f_{T_2}-f_{T_7}\right),\\
%&\alpha_{AZWW,7}=\frac{c_Ws_W}{\Lambda^4}\left(f_{T_1}-f_{T_6}\right),
%\end{split}
%\label{eq.2.8}
%\end{equation}
\begin{equation}
\begin{split}
&\alpha_{2A2Z,0}=\frac{e^2 v^2}{16 \Lambda^4}\frac{1}{c_W s_W}\left(\frac{s_W}{c_W}f_{M_0}+2\frac{c_W}{s_W}f_{M_2}-f_{M_4}\right),\\
&\alpha_{2A2Z,1}=\frac{e^2 v^2}{16 \Lambda^4}\frac{1}{c_W s_W}\left(\frac{s_W}{2c_W}f_{M_7}-\frac{s_W}{c_W}f_{M_1}-2\frac{c_W}{s_W}f_{M_3}-2f_{M_5}\right),\\
&\alpha_{2A2Z,2}=\frac{c_W^2s_W^2}{2\Lambda^4}\left(f_{T_0}+f_{T_1}-2f_{T_6}+4f_{T_8}\right)+\frac{c_W^4+s_W^4}{\Lambda^4}f_{T_5},\\
&\alpha_{2A2Z,3}=\frac{c_W^2s_W^2}{\Lambda^4}\left(f_{T_2}+4f_{T_9}\right)+\frac{\left(c_W^2-s_W^2\right)^2}{2\Lambda^4}f_{T_7},\\
&\alpha_{2A2Z,4}=\frac{c_W^2s_W^2}{\Lambda^4}\left(f_{T_0}+f_{T_1}-2f_{T_5}+4f_{T_8}\right)+\frac{\left(c_W^2-s_W^2\right)^2}{2\Lambda^4}f_{T_6},\\
&\alpha_{2A2Z,5}=\frac{c_W^2s_W^2}{2\Lambda^4}\left(f_{T_2}-2f_{T_7}+4f_{T_9}\right),\\
\end{split}
\label{eq.2.9}
\end{equation}
\begin{equation}
\begin{split}
&\alpha_{A3Z,0}=\frac{e^2 v^2}{8 \Lambda^4}\frac{1}{c_W s_W}\left(f_{M_0}-2f_{M_2}-\frac{c_W^2-s_W^2}{2s_Wc_W}f_{M_4}\right),\\
&\alpha_{A3Z,1}=\frac{e^2 v^2}{8 \Lambda^4}\frac{1}{c_W s_W} \left(2f_{M_3}-f_{M_1}+\frac{f_{M_7}}{2}-\frac{c_W^2-s_W^2}{2s_Wc_W}f_{M_5}\right),\\
&\alpha_{A3Z,2}=\frac{c_W^3s_W}{\Lambda^4}\left(f_{T_0}+f_{T_1}-f_{T_5}-f_{T_6}\right)+\frac{c_Ws_W^3}{\Lambda^4}\left(f_{T_5}+f_{T_6}-4f_{T_8}\right),\\
&\alpha_{A3Z,3}=\frac{c_W^3s_W}{\Lambda^4}\left(f_{T_2}-f_{T_7}\right)+\frac{c_Ws_W^3}{\Lambda^4}\left(f_{T_7}-4f_{T_9}\right),\\
\end{split}
\label{eq.2.10}
\end{equation}
and
\begin{equation}
\begin{split}
&\alpha_{3AZ,0}=\frac{c_W^3s_W}{\Lambda ^4}\left(f_{T_5}+f_{T_6}-4f_{T_8}\right)+\frac{c_Ws_W^3}{\Lambda^4}\left(f_{T_0}+f_{T_1}-f_{T_5}-f_{T_6}\right),\\
&\alpha_{3AZ,1}=\frac{c_W^3s_W}{\Lambda ^4}\left(f_{T_7}-4f_{T_9}\right)+\frac{c_Ws_W^3}{\Lambda ^4}\left(f_{T_2}-f_{T_7}\right).\\
\end{split}
\label{eq.2.11}
\end{equation}

\section{\label{level3}Unitarity bounds}

When aQGCs are taken into account, the cross-section of $VV\to Z\gamma$ grows as $E^4$ which leads to the violation of unitarity~\cite{unitarityHistory1,*unitarityHistory2,*unitarityHistory3}.
It is proposed that~\cite{matchingidea1}, the constraints obtained by experiments should be reported as functions of energy scales, a matching procedure can be built based on this idea~\cite{matchingidea2,*matchingidea3} and has been introduced in the study of aQGCs~\cite{wastudy}.
Such a matching procedure is independent of unitarization methods and can be applied in experiments.
In this paper, we use this method to avoid the violation of unitarity.

The partial wave unitarity bound~\cite{partialwaveunitaritybound,ubnew1,*ubnew2} is used in the above-mentioned matching procedure.
For the subprocess $V_{1,\lambda _1}V_{2,\lambda _2}\to \gamma_{\lambda _3}Z_{\lambda _4}$, where $V_i$ are vector bosons and $\lambda _i$ correspond to the helicities of $V_i$, the amplitude can be expanded as~\cite{partialwaveexpansion}
\begin{equation}
\begin{split}
&\mathcal{M}(V_{1,\lambda _1}V_{2,\lambda _2}\to \gamma_{\lambda _3}Z_{\lambda _4})=8\pi \sum _{J}\left(2J+1\right)\sqrt{1+\delta ^{V_1,\lambda _1}_{V_2,\lambda _2}}e^{i(\lambda-\lambda ') \varphi}d^J_{\lambda \lambda '}(\theta) T^J\\
\end{split}
\label{eq.3.1}
\end{equation}
where $\lambda = \lambda _1-\lambda _2$, $\lambda ' =\lambda _3-\lambda _4$, $\theta$ and $\phi$ are zenith and azimuth angles of the photon in the final state, respectively, and $d^J_{\lambda \lambda '}(\theta)$ are Wigner $d$-functions~\cite{partialwaveexpansion}. The partial wave unitarity bound is $|T^J|\leq 2$~\cite{partialwaveunitaritybound} which is widely used in previous works~\cite{unitarity1,*unitarity2,*unitarity3,*unitarity4}.

The case of applying unitarity bounds to a VBS process is very different from the case that the c.m. energy~(denoted as $\hat{s}$) of the subprocess $VV\to VV$ is determined.
For the latter, imposing unitarity bounds is to assuming that EFT is valid under a fixed energy scale, then there is a maximally allowed coefficient.
One can combine all $VV\to VV$ to give a tightest unitarity bound on a coefficient.
However, for VBS processes, the $\hat{s}$ are not deterministic.
So the unitarity bounds become the maximally allowed energy scales when assuming that EFT is valid.
The allowed energy scale is different for different $VV\to VV$ subprocesses.
Since on a same collider the typical energy scales of $VV\to VV$ subprocesses can be different, we do not combine all $VV\to VV$ subprocesses but combine all $VV\to Z\gamma$ subprocesses to give an upper limit of $\hat{s}$ for a particular final state.

The partial wave expansions of the amplitudes and $|T_J|$ as functions of $\hat{s}$ are listed in Appendix.~\ref{ap}.
For a fixed operator coefficient, $|T_J|\leq 2$ leads to a constraint on $\hat{s}$.
Since we cannot distinguish each vertices strictly in the process $pp\to jj \ell^+\ell ^- \gamma$, we use the strongest bounds, they are
\begin{equation}
\begin{split}
&\hat{s} ^{O_{M_0}}\leq \sqrt{\frac{32\sqrt{2} \pi c_W s_W M_Z^2 \Lambda^4}{|f_{M_0}|e^2 v^2}},\;
 \hat{s} ^{O_{M_1}}\leq \sqrt{\frac{64\sqrt{2}c_Ws_WM_Z^2 \Lambda^4}{|f_{M_1}|e^2v^2}},\\
&\hat{s} ^{O_{M_2}}\leq \sqrt{\frac{16\sqrt{2} \pi c_W s_W M_Z^2 \Lambda^4}{|f_{M_2}|e^2 v^2}},\;
 \hat{s} ^{O_{M_3}}\leq \sqrt{\frac{32\sqrt{2} \pi c_W s_W M_Z^2 \Lambda^4}{|f_{M_3}| e^2   v^2}},\\
&\hat{s} ^{O_{M_4}}\leq \sqrt{\frac{64\sqrt{2}  \pi c_W^2 s_W^2 M_Z^2 \Lambda^4}{|f_{M_4}|(c_W^2-s_W^2)e^2 v^2}},\;
 \hat{s} ^{O_{M_5}}\leq \sqrt{\frac{192 \pi s_W^2 M_WM_Z \Lambda^4}{|f_{M_5}|e^2  v^2}},\\
&\hat{s} ^{O_{M_7}}\leq \sqrt{\frac{128\sqrt{2} \pi c_W s_W M_Z^2 \Lambda^4}{|f_{M_7}| e^2   v^2}},\;
 \hat{s} ^{O_{T_0}}\leq \sqrt{\frac{6\sqrt{2} \pi \Lambda^4}{5|f_{T_0}| c_W^3 s_W}},\\
&\hat{s} ^{O_{T_1}}\leq \sqrt{\frac{6\sqrt{2} \pi \Lambda^4}{5|f_{T_1}| c_W^3 s_W}},\;
 \hat{s} ^{O_{T_2}}\leq \sqrt{\frac{8\sqrt{2} \pi \Lambda^4}{3 |f_{T_2}| c_W^3 s_W}},\\
&\hat{s} ^{O_{T_5}}\leq \sqrt{\frac{\pi \Lambda^4}{|f_{T_5}| c_W^2 s_W^2}},\;
 \hat{s} ^{O_{T_6}}\leq \sqrt{\frac{4\pi \Lambda^4}{|f_{T_6}| (c_W^2 -s_W^2)}},\\
&\hat{s} ^{O_{T_7}}\leq \sqrt{\frac{8\sqrt{2} \pi \Lambda^4}{3 |f_{T_7}| (c_W^2-s_W^2)c_W s_W}},\;
 \hat{s} ^{O_{T_8}}\leq \sqrt{\frac{3 \pi \Lambda^4}{10|f_{T_8}| c_W^2 s_W^2}},\\
&\hat{s} ^{O_{T_9}}\leq \sqrt{\frac{2 \pi \Lambda^4}{3 |f_{T_9}| c_W^2 s_W^2}}.
\end{split}
\label{eq.3.2}
\end{equation}
Taking $O_{M_0}$ as an example, for a given $f_{M_0}/\Lambda ^4$, the results (such as the cross-section) obtained by SMEFT are invalid if $\hat{s}$ is larger then the bound in Eq.~(\ref{eq.3.1}).
Therefore, for a given $f_{M_0}/\Lambda^4$, we use the bound in Eq.~(\ref{eq.3.1}) as an energy cut on the events generated by MC, and compare the signals with the backgrounds under such energy scale.
The results at Z-pole are
\begin{equation}
\begin{split}
&\sqrt{\hat{s} ^{O_{M_0}}}\leq \frac{3.0\Lambda}{|f_{M_0}|^{1/4}}\;{\rm TeV},\;
 \sqrt{\hat{s} ^{O_{M_1}}}\leq \frac{3.6\Lambda}{|f_{M_1}|^{1/4}}\;{\rm TeV},\;
 \sqrt{\hat{s} ^{O_{M_2}}}\leq \frac{2.5\Lambda}{|f_{M_2}|^{1/4}}\;{\rm TeV},\\
&\sqrt{\hat{s} ^{O_{M_3}}}\leq \frac{3.0\Lambda}{|f_{M_3}|^{1/4}}\;{\rm TeV},\;
 \sqrt{\hat{s} ^{O_{M_4}}}\leq \frac{3.4\Lambda}{|f_{M_4}|^{1/4}}\;{\rm TeV},\;
 \sqrt{\hat{s} ^{O_{M_5}}}\leq \frac{3.6\Lambda}{|f_{M_5}|^{1/4}}\;{\rm TeV},\\
&\sqrt{\hat{s} ^{O_{M_7}}}\leq \frac{4.2\Lambda}{|f_{M_7}|^{1/4}}\;{\rm TeV},\;
 \sqrt{\hat{s} ^{O_{T_0}}}\leq \frac{2.0\Lambda}{|f_{T_0}|^{1/4}}\;{\rm TeV},\;
 \sqrt{\hat{s} ^{O_{T_1}}}\leq \frac{2.0\Lambda}{|f_{T_1}|^{1/4}}\;{\rm TeV},\\
&\sqrt{\hat{s} ^{O_{T_2}}}\leq \frac{2.5\Lambda}{|f_{T_2}|^{1/4}}\;{\rm TeV},\;
 \sqrt{\hat{s} ^{O_{T_5}}}\leq \frac{2.1\Lambda}{|f_{T_5}|^{1/4}}\;{\rm TeV},\;
 \sqrt{\hat{s} ^{O_{T_6}}}\leq \frac{2.2\Lambda}{|f_{T_6}|^{1/4}}\;{\rm TeV},\\
&\sqrt{\hat{s} ^{O_{T_7}}}\leq \frac{2.7\Lambda}{|f_{T_7}|^{1/4}}\;{\rm TeV},\;
 \sqrt{\hat{s} ^{O_{T_8}}}\leq \frac{1.5\Lambda}{|f_{T_8}|^{1/4}}\;{\rm TeV},\;
 \sqrt{\hat{s} ^{O_{T_9}}}\leq \frac{1.9\Lambda}{|f_{T_9}|^{1/4}}\;{\rm TeV}.
\end{split}
\label{eq.3.4}
\end{equation}
When $f_X/\Lambda^4\sim 1 \;{\rm TeV}^{-4}$, the constraints are about $\sqrt{\hat{s}}\approx 2\sim 4\;{\rm TeV}$, which can be easily reached at a $13$ or $14$ TeV collider.
As will be shown in Sec.~{\ref{level5}}, the unitarity bound has a significant impact.

\section{\label{level4}Features of the signal}

\begin{figure}[!htbp]
\centering{
\includegraphics[width=0.49\textwidth]{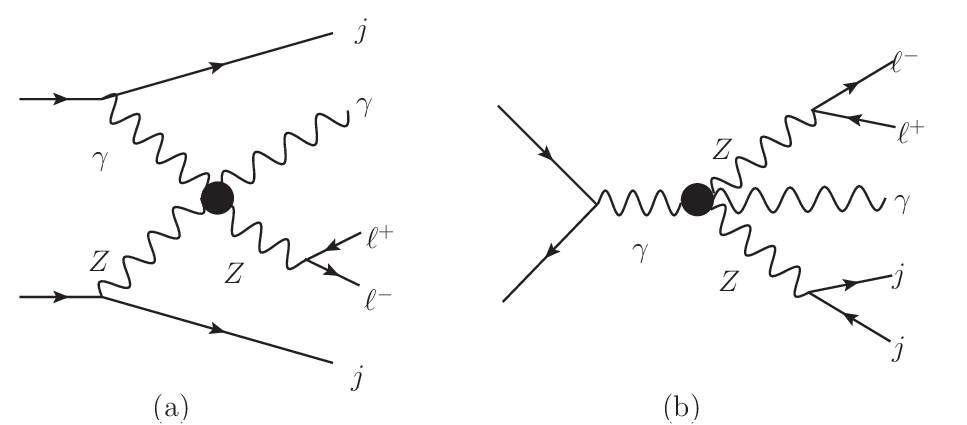}
\caption{\label{Fig:signaldiagram}
Typical Feynman diagrams of aQGC contributions.
}}
\end{figure}
\begin{figure}[!htbp]
\centering{
\includegraphics[width=0.49\textwidth]{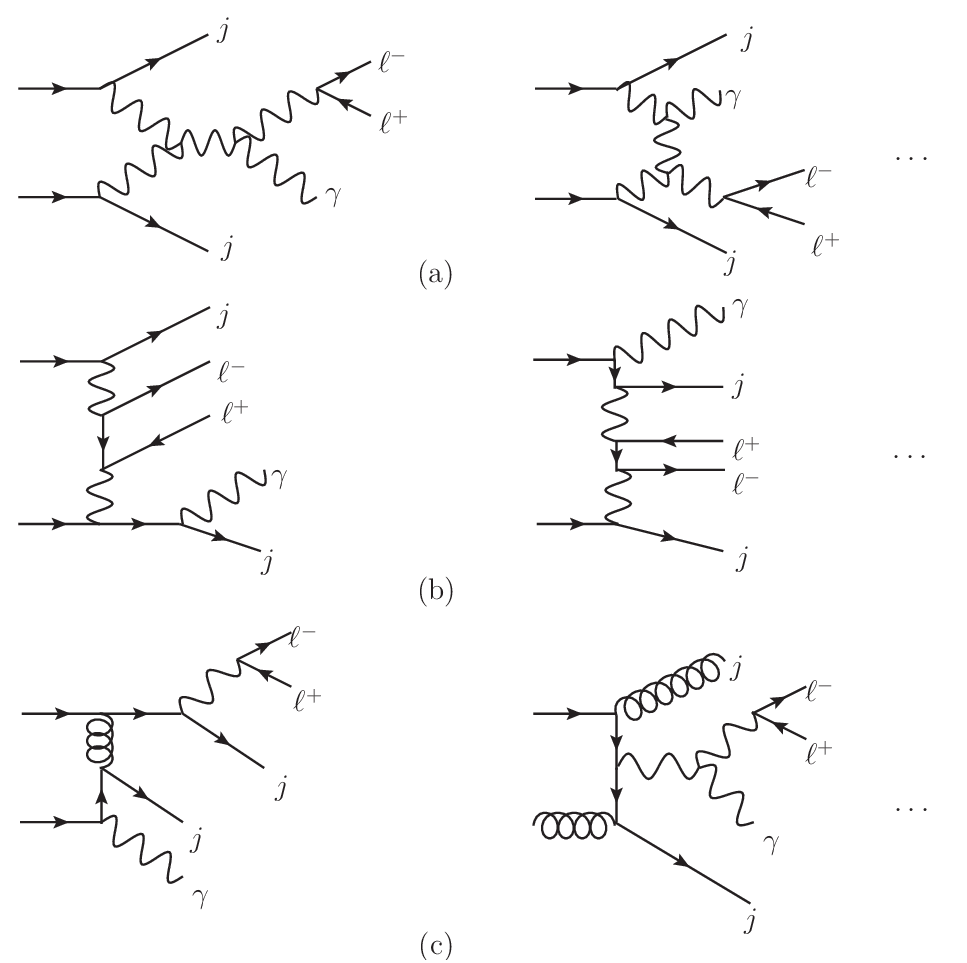}
\caption{\label{Fig:bgdiagram}
Typical Feynman diagrams of the SM backgrounds. which are often categorized as EW-VBS (a), EW-non-VBS (b) and QCD contributions (c).
}}
\end{figure}

The Feynman diagrams of the contribution induced by aQGCs to the process $pp\to jj \ell^+\ell^-\gamma$ are depicted in Fig.~\ref{Fig:signaldiagram}.
The dominant signal is the VBS contribution induced by aQGCs shown in Fig.~\ref{Fig:signaldiagram}.~(a).
The goal of our event selection strategy is to highlight the contribution of Fig.~\ref{Fig:signaldiagram}.~(a).
As shown in Table~\ref{tab.1}, the $O_{M_{0,1,2,3,7}}$ and $O_{T_{0,1,2,8}}$ operators are constraint tightly by previous works, so we concentrate on $O_{M_{4,5}}$ and $O_{T_{5,6,7,9}}$ operators.

The diagrams of the SM backgrounds are shown in Fig.~\ref{Fig:bgdiagram}.
Except for that, there are also other backgrounds due to mistags, such as $pp\to jjj\ell^+\ell^-$ with a jet mistagged as a photon, $pp\to t\bar{t}\gamma \to jj\ell^+\ell^- \gamma \nu\bar{\nu}$ with $b$-jet mistagged, and $pp\to jj ZW^{\pm}\to jj \ell^+\ell^-\ell^{\pm}\nu$ with a lepton mistagged as a photon.
It has been found that the $pp\to jjj\ell^+\ell^-$ background is dominant~\cite{zaexp1,zaexp2}, therefore we only consider this process.

We study the features of the signals and backgrounds using the \verb"MadGraph5_aMC@NLO" toolkit~\cite{madgraph,*feynrules}, we use the basic cuts same as those from the default settings except for $\Delta R_{\ell\ell}$ which is defined as $\sqrt{\Delta \eta_{\ell\ell} ^2+\Delta \phi_{\ell\ell}^2}$ where $\Delta \eta_{\ell\ell}$ and $\Delta \phi_{\ell\ell}$ are the differences between pseudo-rapidities and azimuth angles of leptons, respectively.
As will be explained later, we use $\Delta R_{\ell\ell}>0.2$.
The parton distribution function is \verb"NNPDF2.3"~\cite{NNPDF}.
The events are showered by \verb"PYTHIA8" \cite{pythia}.
A fast detector simulation are performed by \verb"Delphes" \cite{delphes} with the CMS detector card. 
The events of the signals are generated with the largest coefficients listed in Table~\ref{tab.1}, and with one operator at a time.
The analyses of the signal and the background are completed by \verb"MLAnalysis"~\cite{Guo:2023nfu}.  

\begin{figure}[!htbp]
\centering{
\subfigure[$M_{jj}$]{\includegraphics[width=0.32\textwidth]{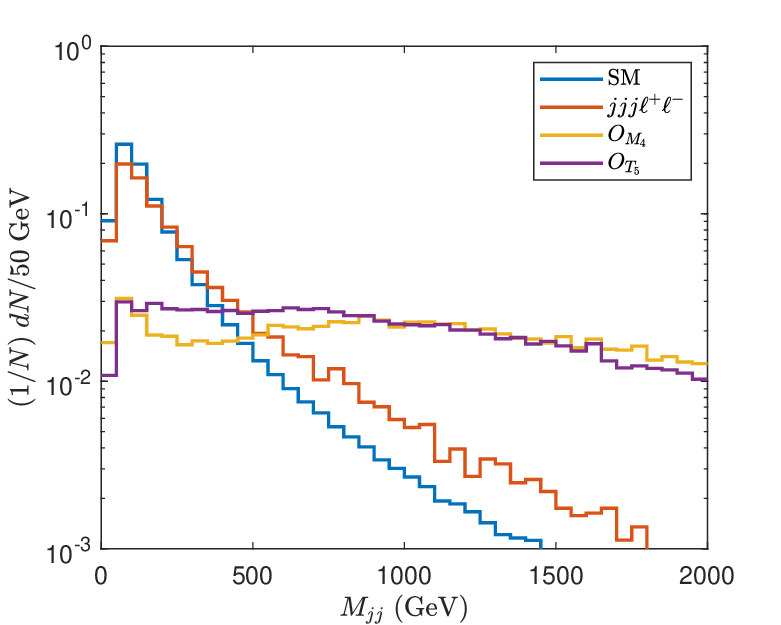}}
\subfigure[$|\Delta y_{jj}|$]{\includegraphics[width=0.32\textwidth]{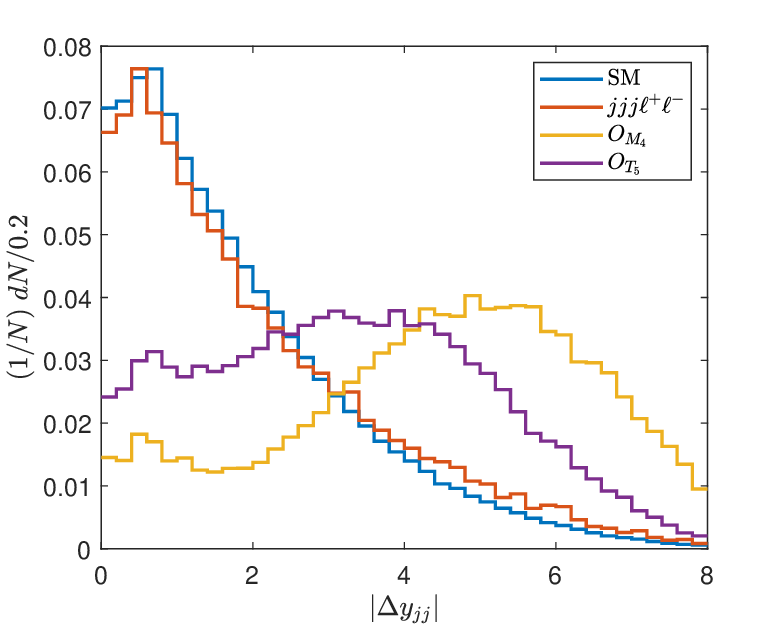}}
\subfigure[$M_{\ell\ell}$]{\includegraphics[width=0.32\textwidth]{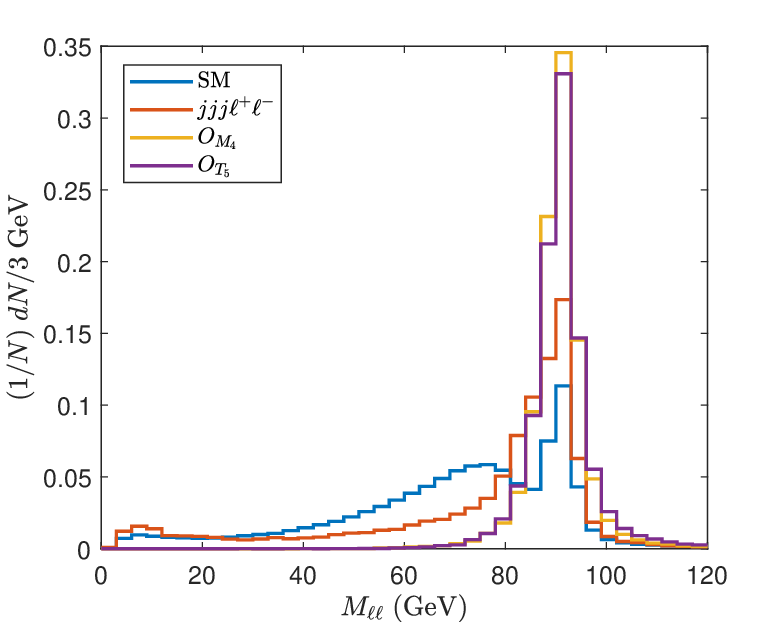}}
\subfigure[$\Delta R _{\ell\ell}$]{\includegraphics[width=0.32\textwidth]{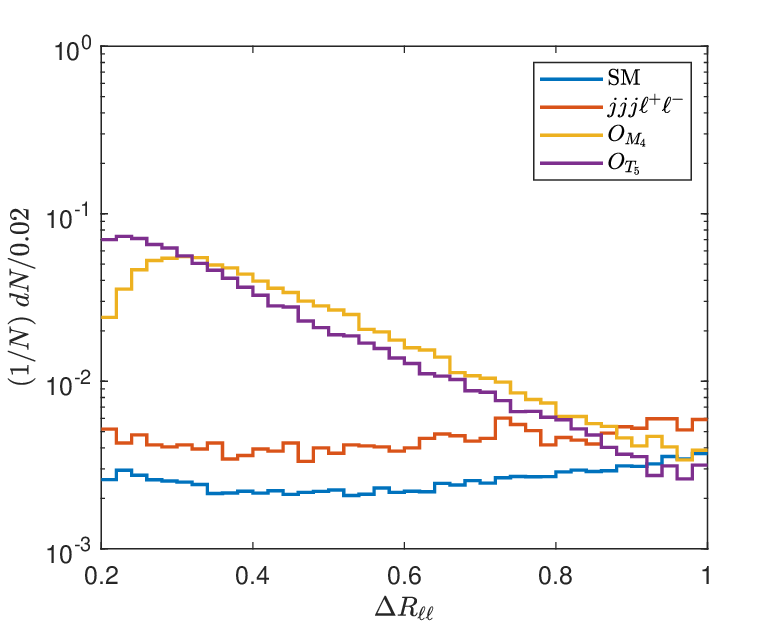}}
\subfigure[$M_{Z\gamma}$]{\includegraphics[width=0.32\textwidth]{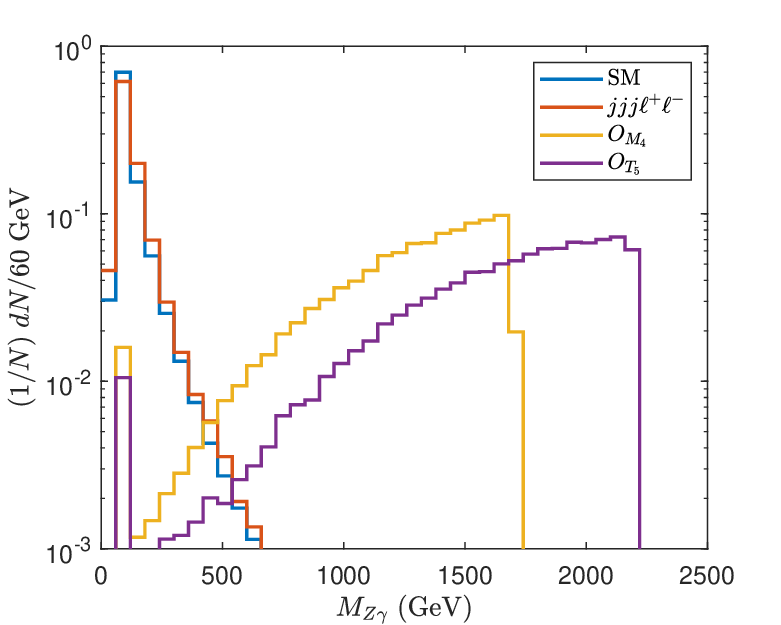}}
\subfigure[$r$]{\includegraphics[width=0.32\textwidth]{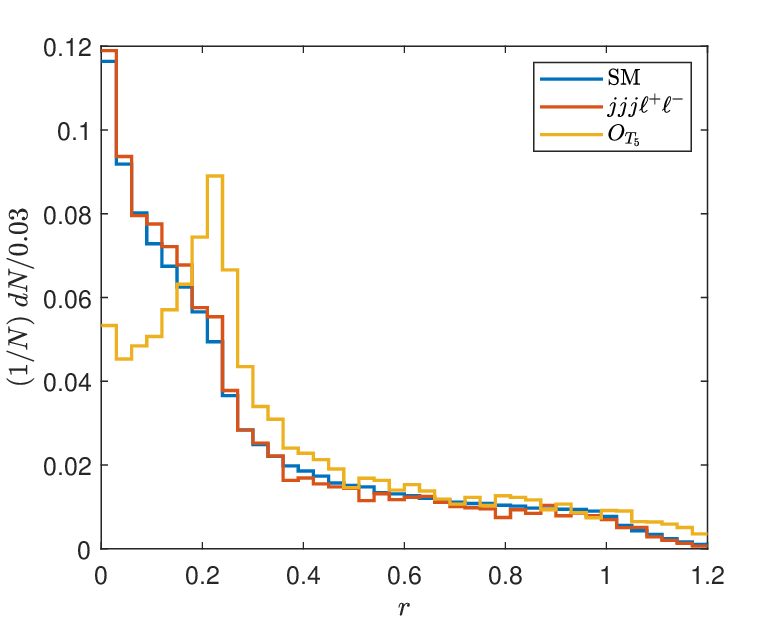}}
\caption{\label{Fig:distributions}The normalized distributions of $M_{jj}$, $|\Delta y_{jj}|$, $M_{\ell\ell}$, $\Delta R _{\ell\ell}$, $M_{Z\gamma}$ and $r$.}}
\end{figure}

First we require the particle numbers in the final state to be $N_j\geq 2$, $N_{\ell^+}\geq 1$, $N_{\ell^-}\geq 1$ and $N_{\gamma}\geq 1$. Apart from that, we require the hardest two leptons to be a lepton and an anti-lepton with same flavor.
The requirement of particle numbers is denoted as $N_{j,\gamma,\ell}$ cut.
In the following of this section, the results are presented after $N_{j,\gamma,\ell}$ cut and unitarity bounds.

Since the dominant signal is the VBS contribution, we use the standard VBS/VBF cut~\cite{vbfcut} which cuts the events with a small $M_{jj}$ or $|\Delta y_{jj}|$ where $M_{jj}$ is the invariant mass of the hardest two jets, and $|\Delta y_{jj}|$ is the difference between the rapidities of the hardest two jets.
We only use $|\Delta y_{jj}|$ for $O_{M_i}$ operators to avoid too many signal events are cut off.
The normalized distributions as functions of $M_{jj}$ and $|\Delta y_{jj}|$ are shown in Figs.~\ref{Fig:distributions}.~(a) and (b).
The distributions for $O_{M_i}$ operators are similar, this is also the case for the $O_{T_i}$ operators, so we only present the distributions of $O_{M_3}$ and $O_{T_5}$ as examples.

In the backgrounds, there are contributions from the diagrams without a $Z$ resonance.
With the approximation that the $Z$ boson is nearly on shell, the invariant mass of the hardest two leptons is approximately $M_{\ell \ell}\approx M_Z$.
$|M_{\ell \ell} -M_Z|$ can be used to exclude the events without a $Z$ resonance.
The distributions as functions of $M_{\ell\ell}$ are shown in Fig.~\ref{Fig:distributions}.~(c).
For the SM backgrounds, both cases with and without $Z$ resonance exist, so there are also peaks at $M_z$.
However, for the signals, the peaks are much sharper because all the diagrams contain a $Z$ resonance.

The contributions from aQGCs grow as $\hat{s}^2$, the signal event typically has a large $\hat{s}$, which can be obtained as $\hat{s}=M^2_{Z\gamma}$, where $M_{Z\gamma}=\sqrt{(p_{\ell^+}+p_{\ell^-}+p_{\gamma})^2}$.
A large $\hat{s}$ also leads to an energetic $Z$ boson, as a consequence, the leptons are close to each other, and a small $\Delta R_{\ell\ell}$ is expected.
The distributions as functions of $\Delta R_{\ell\ell}$ and $M_{Z\gamma}$ are shown in Figs.~\ref{Fig:distributions}.~(d) and (e).
We find that $M_{Z\gamma}$ is a very sensitive variable to discriminate the signals from the backgrounds.

Meanwhile, there is inferred divergence in the backgrounds when leptons are collinear.
On the other hand, $\Delta R_{\ell\ell}$ is related to the isolation of leptons in experiments, therefore a lower bound of $\Delta R_{\ell\ell}$ has to be set.
A smaller lower bound of $\Delta R_{\ell\ell}$ is helpful to preserve the signal events.
In experiments, $\Delta R_{\ell\ell}$ can be as small as $0.2$~\cite{zaexp1}, so we use $\Delta R_{\ell\ell}>0.2$ in the basic cuts.
One shall be careful that the contribution from aQGCs can not be obtained as $\sigma (pp\to jjZ\gamma)$ multiplied by $Br(Z\to\ell ^+\ell^-)$, because about $60\%$ of signal events are missed due to this $\Delta R_{\ell\ell}$ cut.

\begin{figure}[!htbp]
\centering{
\subfigure[SM]{\includegraphics[width=0.48\textwidth]{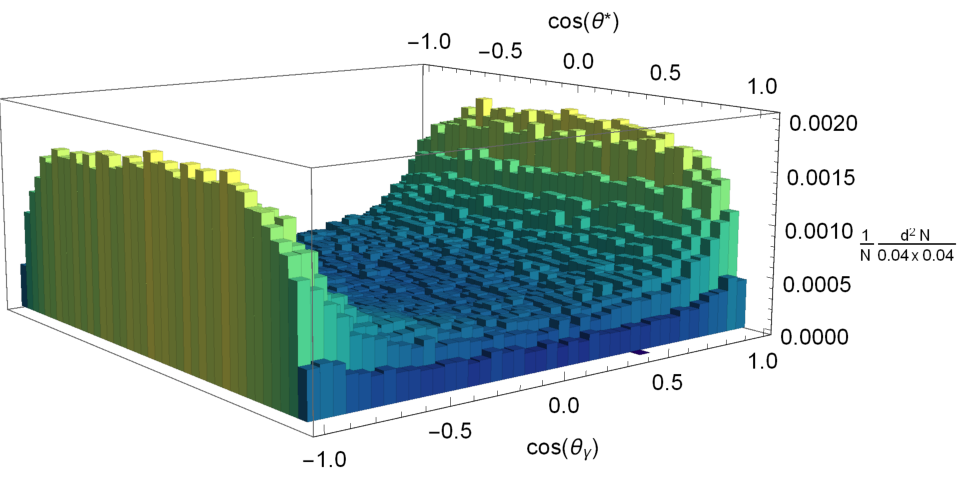}}
\subfigure[$jjj\ell^+\ell^-$]{\includegraphics[width=0.48\textwidth]{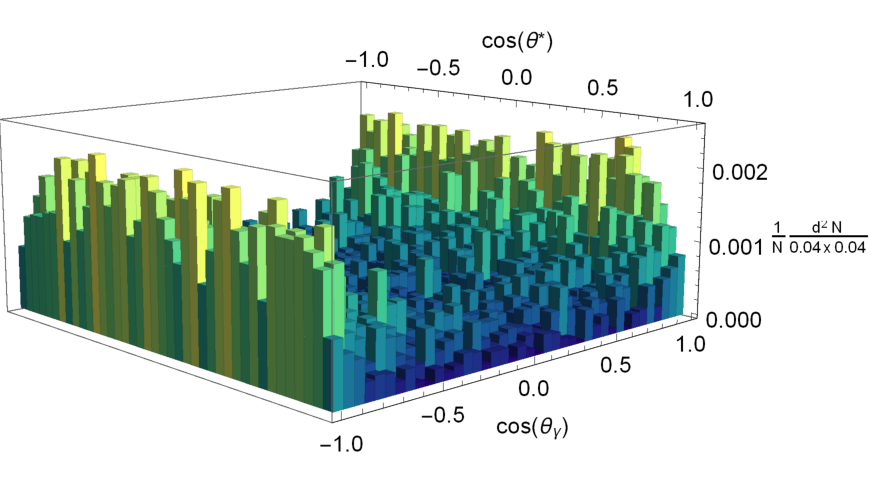}}
\subfigure[$O_{M_4}$]{\includegraphics[width=0.48\textwidth]{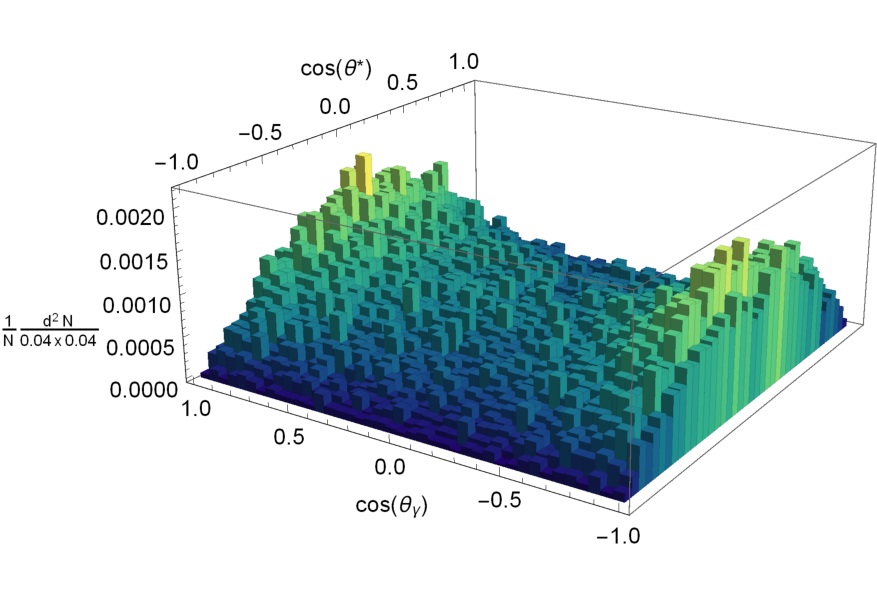}}
\subfigure[$O_{T_5}$]{\includegraphics[width=0.48\textwidth]{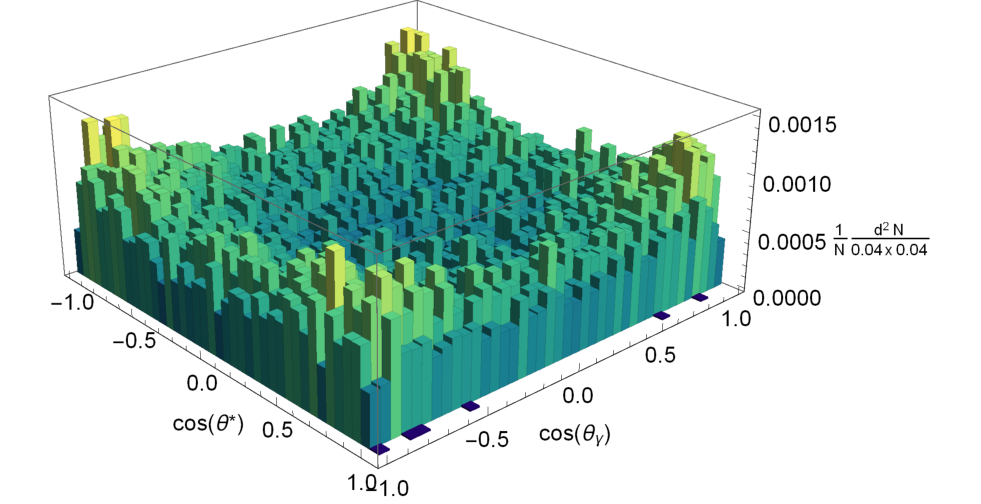}}
\caption{\label{Fig:distribution3d}
The normalized distributions of $\cos (\theta_{\gamma})$ and $\cos (\theta ')$. Each bin corresponds to $d \cos (\theta_{\gamma}) \times d \left(\cos \theta'\right)=0.04\times 0.04$~($50\times 50$ bins).}}
\end{figure}

The polarization properties and resulting angular distributions have been used to distinguish signals from backgrounds~\cite{wastudy,zzpolarization,*gggpolarization,*dijetgpolarization}.
For $O_{T_i}$ operators, we find that the $Z$ bosons are dominantly transversely polarized from the partial wave expansions\footnote{The partial wave expansions are listed in Appendix.~\ref{ap}}.
This leads to a unique angular distribution of the lepton in the $Z$ boson helicity frame.
Denoting $\theta '$ as the zenith angle of lepton in the $Z$ boson helicity frame, similar as Ref.~\cite{wastudy}, we investigate the correlation between $\cos(\theta ')$ and $\cos (\theta _{\gamma})$ where $\theta _{\gamma}$ is the zenith angle of photon in the rest frame of $pp$.
The normalized distributions of $\cos(\theta ')$ and $\cos (\theta _{\gamma})$ are shown in Fig.~\ref{Fig:distribution3d}.
The distributions of the backgrounds peak at a small $\cos(\theta ')$ and a large $|\cos (\theta _{\gamma})|$, therefore we introduce observable $r$ defined in Ref.~\cite{wastudy} to identify the signals of $O_{T_i}$ operators, except that we replace the $L_p$~(which is an approximation of $(\cos(\theta ') + 1)/ 2$~\cite{lp}) with $(\cos(\theta ') + 1)/ 2$,
\begin{equation}
\begin{split}
&r=\left(1-\left|\cos(\theta _{\gamma})\right|\right)^2+\frac{\cos (\theta ')^2}{4}.
\end{split}
\label{eq.4.1}
\end{equation}
The distributions as functions of $r$ are shown in Fig.~\ref{Fig:distributions}.~(f).
It was pointed out that the correlation between $r$ and $\hat{s}$ is small~\cite{wastudy}, thus the cut using $r$ can serve as a complementary to the cut using $\hat{s}$.

The event selection strategy proposed for aQGCs and its effects are listed in Table~\ref{Tab:cutflow}.
The large SM backgrounds especially the $jjj\ell^+\ell^-$ background can be effectively reduced by our selection strategy.

%\begin{table*}
%\begin{center}
%\begin{tabular}{c|c|c|c|c|c|c|c|c|c}
%\hline
% & SM & $jjj\ell^+\ell^-$ & $O_{M_3}$ & $O_{M_4}$ & $O_{M_5}$ & $O_{T_5}$ & $O_{T_6}$ & $O_{T_7}$ & $O_{T_9}$ \\
%\hline
%$N_{j,\gamma,\ell}$ cut & $728.3$ & $791.9$ & $1.48$ & $0.695$ & $2.39$ & $0.883$ & $1.04$ & $0.916$ & $2.10$\\
%\hline
%unitarity bounds &  &  & $0.208$ & $0.160$ & $0.530$ & $0.247$ & $0.234$ & $0.232$ & $0.302$\\
%\hline
%$M_{jj}>100$ GeV & $591.3$ & $677.1$ & $0.202$ & $0.156$ & $0.514$ & $0.242$ & $0.229$ & $0.227$ & $0.294$ \\
%\hline
%$\Delta R_{\ell\ell}<0.8$ & $44.3$ & $90.5$ & $0.168$ & $0.133$ & $0.448$ & $0.212$ & $0.193$ & $0.197$ & $0.245$ \\
%\hline
%$|m_{\ell\ell}-M_Z|<25$ GeV & $9.00$ & $37.8$ & $0.166$ & $0.131$ & $0.443$ & $0.210$ & $0.191$ & $0.195$ & $0.243$ \\
%\hline
%$M_{Z\gamma}>1$ TeV & $0.173$ & $0.044$ & $0.119$ & $0.110$ & $0.365$ & $0.198$ & $0.170$ & $0.180$ & $0.211$ \\
%\hline
%$|\Delta y_{jj}|>0.8$& $0.140$ & $0.044$ & $0.114$ & $0.105$ & $0.349$ &  &  &  &  \\
%$r>0.08$ & $0.106$ & $0.011$ &  &  &  & $0.168$ & $0.140$ & $0.152$ & $0.176$ \\
%\hline
%\end{tabular}
%\end{center}
%\caption{\label{Tab:cutflow}The cross-sections~(fb) of the signals and backgrounds after unitarity bounds and cuts.}
%\end{table*}

\begin{table*}
\begin{center}
\begin{tabular}{c|c|c|c|c|c|c|c|c}
\hline
 & SM & $jjj\ell^+\ell^-$ & $O_{M_4}$ & $O_{M_5}$ & $O_{T_5}$ & $O_{T_6}$ & $O_{T_7}$ & $O_{T_9}$ \\
\hline
$N_{j,\gamma,\ell}$ cut & $728.3$ & $791.9$ & $0.695$ & $2.39$ & $0.883$ & $1.04$ & $0.916$ & $2.10$\\
\hline
unitarity bounds &  &  & $0.208$ & $0.160$ & $0.247$ & $0.234$ & $0.232$ & $0.302$\\
\hline
$M_{jj}>100$ GeV & $591.3$ & $677.1$ & $0.156$ & $0.514$ & $0.242$ & $0.229$ & $0.227$ & $0.294$ \\
\hline
$\Delta R_{\ell\ell}<0.8$ & $44.3$ & $90.5$ & $0.133$ & $0.448$ & $0.212$ & $0.193$ & $0.197$ & $0.245$ \\
\hline
$|m_{\ell\ell}-M_Z|<25$ GeV & $9.00$ & $37.8$ & $0.131$ & $0.443$ & $0.210$ & $0.191$ & $0.195$ & $0.243$ \\
\hline
$M_{Z\gamma}>1$ TeV & $0.173$ & $0.044$ & $0.110$ & $0.365$ & $0.198$ & $0.170$ & $0.180$ & $0.211$ \\
\hline
$|\Delta y_{jj}|>0.8$& $0.140$ & $0.044$ & $0.105$ & $0.349$ &  &  &  &  \\
$r>0.08$ & $0.106$ & $0.011$ &  &  &  $0.168$ & $0.140$ & $0.152$ & $0.176$ \\
\hline
\end{tabular}
\end{center}
\caption{\label{Tab:cutflow}The cross-sections~(fb) of the signals and backgrounds after unitarity bounds and cuts.}
\end{table*}

\section{\label{level5}Numerical results}

\begin{figure}[!htbp]
\centering{
\subfigure{\includegraphics[width=0.48\textwidth]{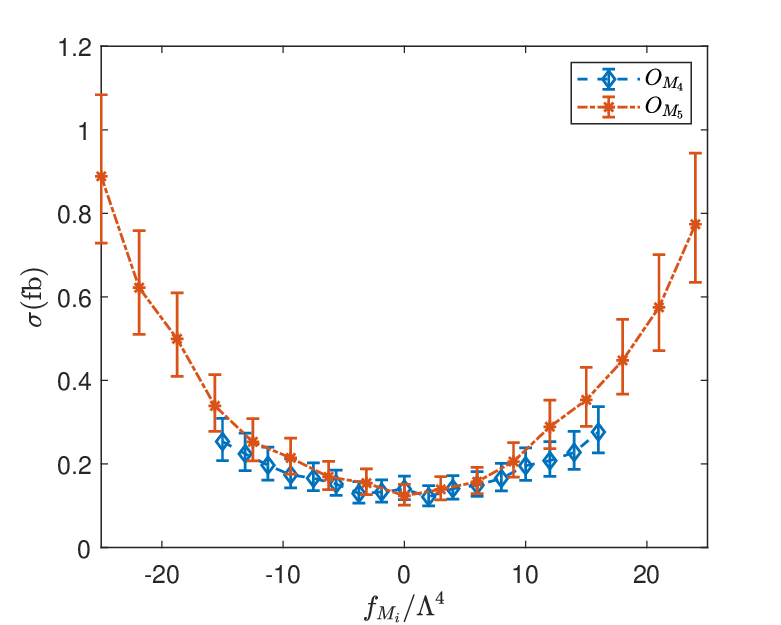}}
\subfigure{\includegraphics[width=0.48\textwidth]{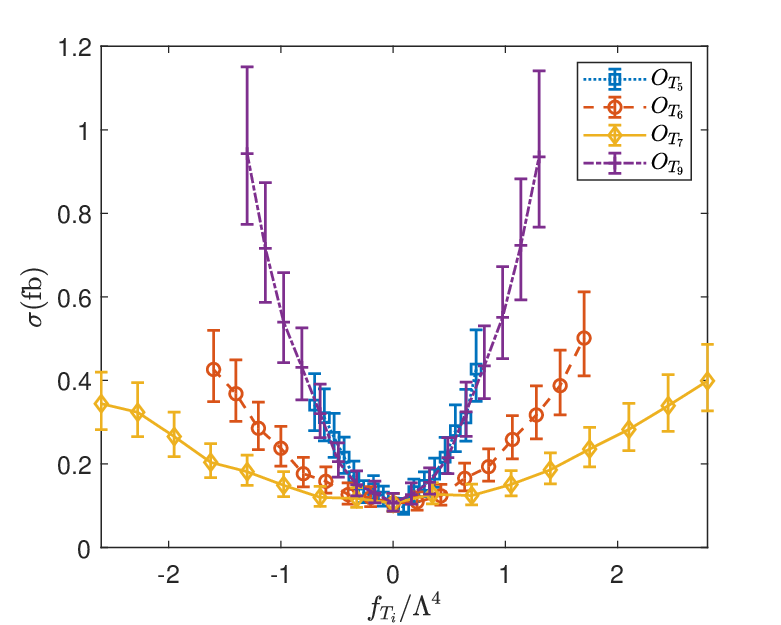}}
\caption{\label{Fig:cs}
The cross-sections as functions of $f_X/\Lambda^4$ without unitarity bounds.}}
\end{figure}

\begin{figure}[!htbp]
\centering{
\subfigure{\includegraphics[width=0.48\textwidth]{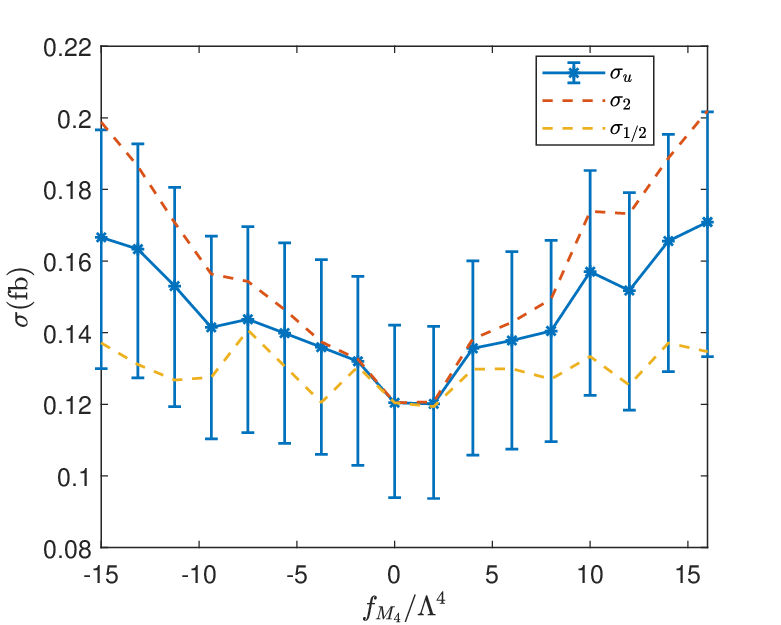}}
\subfigure{\includegraphics[width=0.48\textwidth]{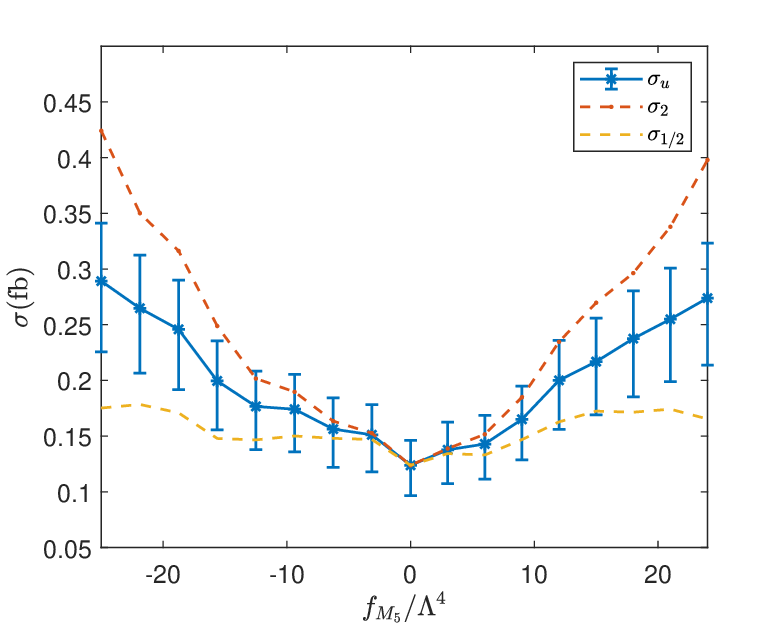}}\\
\subfigure{\includegraphics[width=0.48\textwidth]{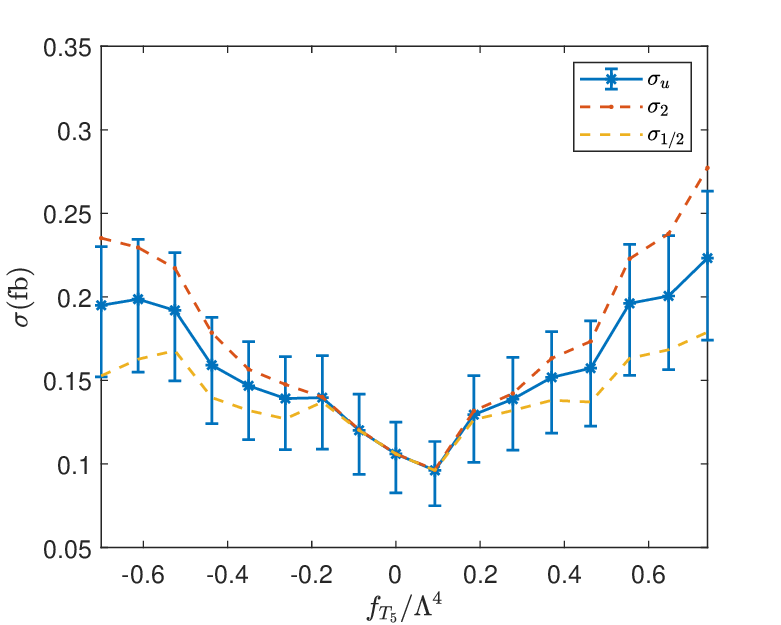}}
\subfigure{\includegraphics[width=0.48\textwidth]{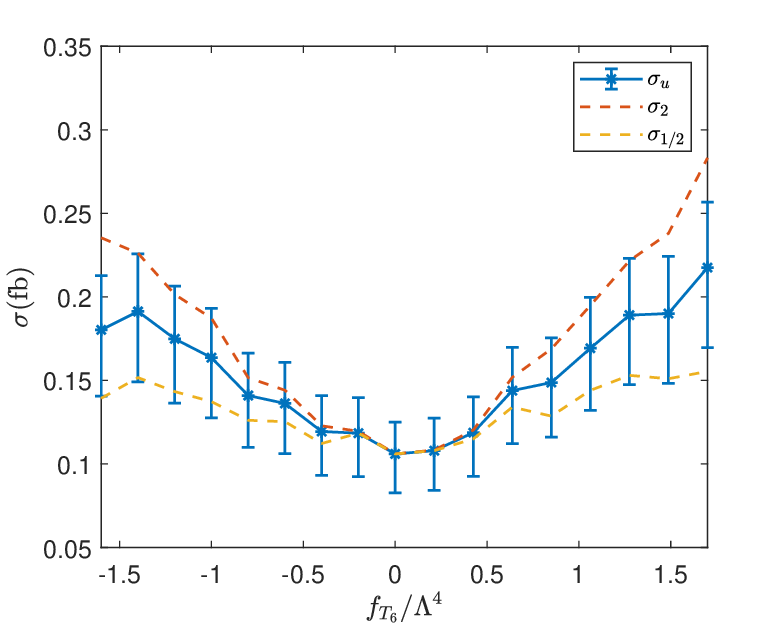}}\\
\subfigure{\includegraphics[width=0.48\textwidth]{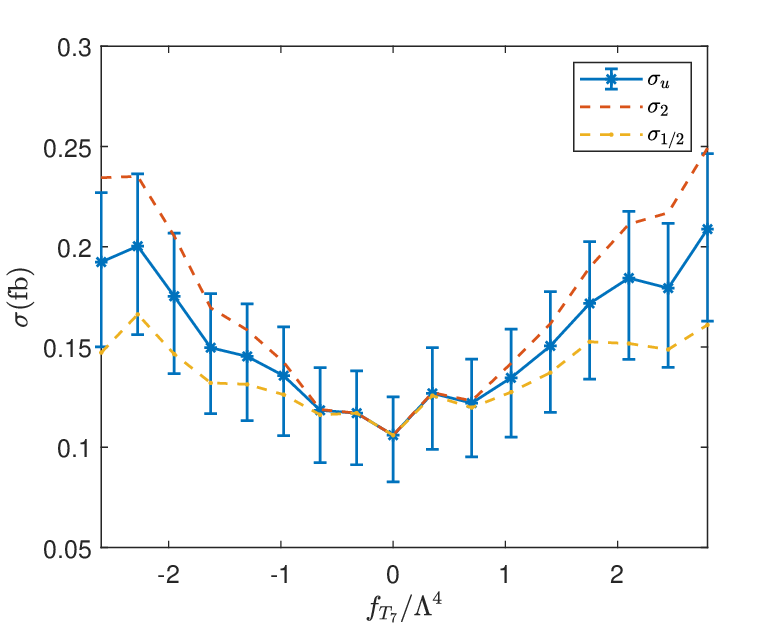}}
\subfigure{\includegraphics[width=0.48\textwidth]{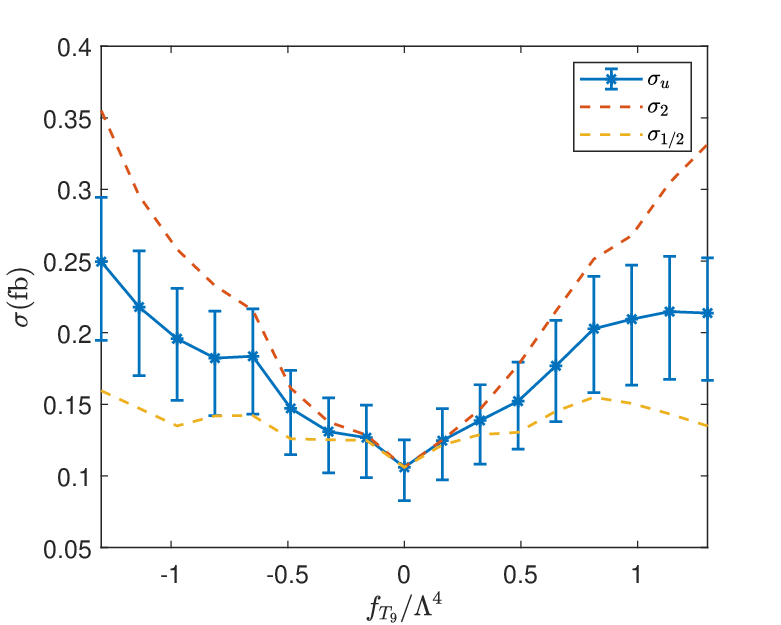}}\\
\caption{\label{Fig:csu}
The cross-sections as functions of $f_X/\Lambda^4$ with unitarity bounds.}}
\end{figure}

By scanning the parameter space listed in Table~\ref{tab.1}, the results are obtained with one operator at a time.
The triboson channel induced by aQGCs, and all possible interferences are included.

To illustrate the effects of unitarity bounds, we give both the results without and with unitarity bounds.
It has been pointed out that unitarity limits depend on hypothesis on the number of operators contributing to aQCGs~\cite{ubnew}.
To study the robustness of the numerical results after unitarity bounds, we present the cross-sections by varying the energy cuts on the $\hat{s}^2$ by factors $1/2$ and $2$.
The cross-sections without unitarity bounds, with unitarity bounds, with the unitarity bounds varied by factors $1/2$ and $2$ are denoted as $\sigma $, $\sigma _ u$, $\sigma _ {1/2}$ and $\sigma _ 2$, respectively.

$\sigma$ for $O_{M_{4,5}}$ and $O_{T_{5,6,7,9}}$ are shown in Fig.~\ref{Fig:cs}.
The statistical errors are negligible compared with the systematic errors which are shown as errorbars.
The cross-sections are bilinear functions of coefficients without the unitarity bounds.
It can be seen that the numerical results fit the bilinear functions well.
The symmetry axes of the parabolas are close to $0$, indicating that the interference between aQGCs and the SM is negligible within the range of coefficients we used.

With unitarity bounds applied, the cross-sections are almost linear functions of the coefficients.
For the largest coefficients, the cross-sections are suppressed by about an order of magnitude for $O_{M_5}$ and $O_{T_9}$.
Such a significant suppression indicates the necessity of unitarity bounds.
It can also be seen from $\sigma _{1/2}$ and $\sigma _2$ that the specific values of unitarity bounds have smaller impacts when the coefficients are smaller.
For the largest coefficients we are using, the difference among $\sigma _u$, $\sigma _{1/2}$ and $\sigma _2$ are at the same order as the systematic errors estimated by varying the QCD scales.
However, when the coefficients are tightened by another order of magnitude, the variations in unitarity bounds will no longer be important compared to systematic errors.

The sensitivities to the operators are estimated by using statistical significance defined as $\mathcal{S}_{stat}\equiv N_S/\sqrt{N_S+N_B}$, where $N_S$ is the number of signal events, and $N_B$ is the number of the background events\footnote{The $jjj\ell^+\ell^-$ production is also included in the background events}.
For each coefficient used to generate the events, the unitarity bound is set accordingly, afterwards $\mathcal{S}_{stat}$ can be obtained.
The expected constraints when NP was not observed are estimated as the lowest positive coefficient and greatest negative coefficient with $\mathcal{S}_{stat}>2$.
The numerical results are shown in Table~\ref{Tab:coefficients1} for luminosity $\mathcal{L}=300\;{\rm fb}^{-1}$ and $3\;{\rm ab}^{-1}$~\cite{HLLHC}.
We find that, at $\sqrt{s}=14$ TeV the constraints cannot be significantly strengthened due to the suppressions of unitarity bounds.
Without unitarity bounds, the signals of aQGCs grows as $E^4$.
However, if new particles do not appear as the energy growing, then increasing the luminosity is as important as increasing the energy to search for the signal of NP.

\begin{table}
\begin{center}
\begin{tabular}{c|c|c}
\hline
 & $300\;{\rm fb}^{-1}$ & $3\;{\rm ab}^{-1}$ \\
\hline
%$f_{M_3}/\Lambda^4$ & $[-18.4, 15.8]$ & $[-7.9, 5.3]$\\
$f_{M_4}/\Lambda^4$ & $[-15.0, 16.0]$ & $[-1.8, 4.0]$ \\
$f_{M_5}/\Lambda^4$ & $[-12.5, 10.0]$ & $[-3.0, 4.0]$ \\
$f_{T_5}/\Lambda^4$ & $[-0.40, 0.37]$ & $[-0.09, 0.15]$\\
$f_{T_6}/\Lambda^4$ & $[-1.0, 0.9]$ & $[-0.4, 0.43]$\\
$f_{T_7}/\Lambda^4$ & $[-1.7, 1.4]$ & $[-0.7, 0.7]$\\
$f_{T_9}/\Lambda^4$ & $[-0.55, 0.50]$ & $[-0.15, 0.15]$\\
\hline
\end{tabular}
\end{center}
\caption{\label{Tab:coefficients1}The constraints (${\rm TeV}^{-4}$) on the operators at $14\;{\rm TeV}$ with $\mathcal{L}=300\; {\rm fb^{-1}}$ and $\mathcal{L}=3\; {\rm ab^{-1}}$ when $\mathcal{S}_{stat}>2$.}
\end{table}

\section{\label{level6}Summary}

In this paper, we study the signals of aQGCs in the process $pp\to jj\ell^+\ell^-\gamma$.
To avoid the violation of unitarity, we use a matching procedure to impose unitarity bounds.
We compare the cross-sections with and without aQGCs under a certain energy scale such that the unitarity is not violated.
This method is independent of unitrization schemes and can be used in experiments.
According to the numerical results at $14$ TeV, the contributions of aQGCs are suppressed significantly by unitarity bounds.
For $O_{M_i}$ operators, the signals can be suppressed by almost one order of magnitude.
Such a significant suppression indicates that there are many events generated in the region that the SMEFT is not valid, and the unitarity bounds are necessary.

The event selection strategy specifically for the aQGCs in the process $pp\to jj\ell^+\ell^-\gamma$ is investigated.
We find that the most sensitive observable to distinguish the signals from backgrounds is $\hat{s}$.
The angular distribution induced by the polarization effect of aQGCs is also studied.
Another observable $r$ originating from the polarization feature of the $Z$ boson is shown to be effective in discriminating the signals induced by $O_{T_i}$ operators.
The sensitivity estimates on the operators at $14$ TeV $300\;{\rm fb}^{-1}$ and $3\;{\rm ab}^{-1}$ are obtained.
We find that it is important to increase the luminosity if new particles are absent as the energy increases.
The constraints from the process $pp\to jj\ell^+\ell^-\gamma$ can contribute to the combined limits.

\section*{ACKNOWLEDGMENT}

\noindent
This work was supported in part by the National Natural Science Foundation of China under Grants No.11905093, 11875157 and 11947402; and by the Doctoral Start-up Foundation of Liaoning Province No.2019-BS-154.

\appendix

\section{\label{ap}Unitarity bounds}

\subsection{\label{sec:azwwvertexu}\texorpdfstring{Unitarity bounds from $WW \to \gamma Z$ process}{Unitarity bounds from WW to AZ process}}

\begin{table*}
\begin{center}
\begin{tabular}{c|c|c}
\hline
Process &  Partial wave expansion & max $|T_J|$ \\
\hline
$\mathcal{M}^{(O_{M_0})}(W_0W_0\to \gamma_+Z_+)$ & $-\frac{f_{M_0}}{\Lambda^4}d_{0,0}^0\frac{c_W e^2 v^2}{2M_W^2 s_W}\hat{s}^2$ & $\frac{\hat{s}^2 |f_{M_0}| c_W e^2 v^2}{16\sqrt{2}\pi M_W^2 s_W \Lambda^4}$ \\
\hline
$\mathcal{M}^{(O_{M_1})}(W_0W_0\to \gamma_+Z_+)$ & $\frac{f_{M_1}}{\Lambda^4}d_{0,0}^0\frac{c_W e^2 v^2}{8M_W^2 s_W}\hat{s}^2$ & $\frac{\hat{s}^2 |f_{M_1}|c_W e^2 v^2}{64\sqrt{2}\pi M_W^2 s_W \Lambda^4}$ \\
\hline
$\mathcal{M}^{(O_{M_4})}(W_0W_0\to \gamma_+Z_+)$ & $-\frac{f_{M_4}}{\Lambda^4}d_{0,0}^0\frac{ e^2 v^2 (c_W^2-s_W^2)}{4 s_W^2 M_W^2} \hat{s}^2$ & $\frac{\hat{s}^2 |f_{M_4}| e^2 v^2 (c_W^2-s_W^2)}{32\sqrt{2}\pi s_W^2 M_W^2 \Lambda^4}$ \\
\hline
$\mathcal{M}^{(O_{M_5})}(W_+W_0\to \gamma_+Z_0)$ & $\frac{f_{M_5}}{\Lambda^4}d_{1,1}^1\frac{e^2  v^2}{4s_W^2 M_WM_Z} \hat{s}^2$ & $\frac{\hat{s}^2 |f_{M_5}|e^2  v^2}{96\pi s_W^2 M_WM_Z \Lambda^4}$\\
\hline
$\mathcal{M}^{(O_{M_7})}(W_0W_0\to \gamma_+Z_+)$ & $-\frac{f_{M_7}}{\Lambda^4}\frac{e^2 c_Wv^2}{16 s_W M_W^2}\hat{s}^2$ & $\frac{\hat{s}^2|f_{M_7}|e^2 c_Wv^2}{128\sqrt{2}\pi s_W M_W^2 \Lambda^4}$\\
\hline
$\mathcal{M}^{(O_{T_0})}(W_+W_+\to \gamma_+Z_+)$ & $8 c_W s_W \frac{f_{T_0}}{\Lambda^4}d_{0,0}^0\hat{s}^2$  & $\frac{\hat{s}^2 |f_{T_0}| c_W s_W}{\sqrt{2}\pi \Lambda^4}$\\
\hline
$\mathcal{M}^{(O_{T_1})}(W_+W_+\to \gamma_-Z_-)$ & $\left(\frac{8}{3}d_{0,0}^0 + \frac{4}{3}d_{0,0}^2\right)\frac{f_{T_1}}{\Lambda^4}c_W s_W \hat{s}^2$ & $\frac{\hat{s}^2|f_{T_1}|c_W s_W }{3\sqrt{2}\pi \Lambda^4}$\\
\hline
$\mathcal{M}^{(f_{T_2})}(W_+W_+\to \gamma_+Z_+)$ & $2\frac{f_{T_2}}{\Lambda^4}c_W s_W d_{0,0}^0\hat{s}^2$ & $\frac{\hat{s}^2 |f_{T_2}|c_W s_W }{4\sqrt{2}\pi \Lambda^4}$ \\
\hline
\end{tabular}
\end{center}
\caption{\label{tab.wwazu}The partial-wave expansions of the $WW\to Z\gamma$ amplitudes at the leading order. $\theta$ and $\varphi$ are zenith and azimuth angles of $\gamma$  in the final state.}
\end{table*}
The partial wave expansion of the $WW \to \gamma Z$ amplitudes are shown in Table~\ref{tab.wwazu}.
For each operator, only those amplitudes with the largest $|T_J|$ are shown.
For those operators not listed in Table~\ref{tab.wwazu}, the amplitudes can be obtained by using
\begin{equation}
\begin{split}
&\mathcal{M}^{(O_{M_2})}(WW\to \gamma Z)=-2\frac{f_{M_2}}{f_{M_0}}\mathcal{M}^{(O_{M_0})}(WW\to \gamma Z),\\
&\mathcal{M}^{(O_{M_3})}(WW\to \gamma Z)=-2\frac{f_{M_3}}{f_{M_1}}\mathcal{M}^{(O_{M_1})}(WW\to \gamma Z),\\
&\mathcal{M}^{(O_{T_5})}(WW\to \gamma Z)=-\frac{f_{T_5}}{f_{T_0}}\mathcal{M}^{(O_{T_0})}(WW\to \gamma Z),\\
&\mathcal{M}^{(O_{T_6})}(WW\to \gamma Z)=-\frac{f_{T_6}}{f_{T_1}}\mathcal{M}^{(O_{T_1})}(WW\to \gamma Z),\\
&\mathcal{M}^{(O_{T_7})}(WW\to \gamma Z)=-\frac{f_{T_7}}{f_{T_2}}\mathcal{M}^{(O_{T_2})}(WW\to \gamma Z).\\
\end{split}
\label{eq.a.1}
\end{equation}
The unitarity bounds are
\begin{equation}
\begin{split}
&\hat{s} ^{O_{M_0}}\leq \sqrt{\frac{32\sqrt{2} \pi M_W^2 s_W \Lambda^4}{|f_{M_0}| c_W e^2 v^2}},\;
 \hat{s} ^{O_{M_1}}\leq \sqrt{\frac{128\sqrt{2} \pi M_W^2 s_W \Lambda^4}{|f_{M_1}| c_W e^2 v^2}},\\
&\hat{s} ^{O_{M_2}}\leq \sqrt{\frac{16\sqrt{2} \pi M_W^2 s_W \Lambda^4}{|f_{M_2}| c_W e^2 v^2}},\;
 \hat{s} ^{O_{M_3}}\leq \sqrt{\frac{64\sqrt{2} \pi M_W^2 s_W \Lambda^4}{|f_{M_3}| c_W e^2 v^2}},\\
&\hat{s} ^{O_{M_4}}\leq \sqrt{\frac{64\sqrt{2} \pi s_W^2 M_W^2 \Lambda^4}{|f_{M_4}| e^2 v^2 (c_W^2-s_W^2)}},\\
&\hat{s} ^{O_{M_5}}\leq \sqrt{\frac{192 \pi s_W^2 M_WM_Z \Lambda^4}{|f_{M_5}|e^2  v^2}},\;
 \hat{s} ^{O_{M_7}}\leq \sqrt{\frac{256\sqrt{2} \pi s_W M_W^2 \Lambda^4}{|f_{M_7}|e^2 c_Wv^2}},\\
&\hat{s} ^{O_{T_0}}\leq \sqrt{\frac{2\sqrt{2} \pi \Lambda^4}{|f_{T_0}| c_W s_W}},\;
 \hat{s} ^{O_{T_1}}\leq \sqrt{\frac{6\sqrt{2} \pi \Lambda^4}{|f_{T_1}|c_W s_W}},\\
&\hat{s} ^{O_{T_2}}\leq \sqrt{\frac{8\sqrt{2} \pi \Lambda^4}{|f_{T_2}|c_W s_W}},\;
 \hat{s} ^{O_{T_5}}\leq \sqrt{\frac{2\sqrt{2} \pi \Lambda^4}{|f_{T_5}| c_W s_W}},\\
&\hat{s} ^{O_{T_6}}\leq \sqrt{\frac{6\sqrt{2} \pi \Lambda^4}{|f_{T_6}|c_W s_W}},\;
 \hat{s} ^{O_{T_7}}\leq \sqrt{\frac{8\sqrt{2} \pi \Lambda^4}{|f_{T_7}|c_W s_W}}.
\end{split}
\label{eq.a.2}
\end{equation}

\subsection{\label{sec:aazzvertexu}\texorpdfstring{Unitarity bounds from $\gamma Z \to \gamma Z$ process}{Unitarity bounds from AZ to AZ process}}

\begin{table*}
\begin{center}
\begin{tabular}{c|c|c}
\hline
Process &  Partial wave expansion & max $|T_J|$ \\
\hline
$\mathcal{M}^{(f_{M_0})}(\gamma_-Z_0\to \gamma_+Z_0)$ & $\frac{f_{M_0}}{\Lambda^4}e^{-2i \phi}\left(-\frac{3}{2}d_{-1,1}^1 + \frac{1}{2}d_{-1,1}^2\right)\frac{e^2  v^2 }{4 c_W^2 M_Z^2}\hat{s}^2$ & $\frac{ \hat{s}^2 |f_{M_0}| e^2  v^2 }{64\pi c_W^2 M_Z^2 \Lambda^4}$\\
\hline
$\mathcal{M}^{(f_{M_1})}(\gamma_+Z_0\to \gamma_+Z_0)$ & $\frac{f_{M_1}}{\Lambda^4}d_{1,1}^1\frac{e^2  v^2 }{4 c_W^2 M_Z^2 }\hat{s}^2$ & $\frac{\hat{s}^2|f_{M_1}| e^2   v^2 }{96\pi c_W^2 M_Z^2 \Lambda^4}$\\
\hline
$\mathcal{M}^{(f_{T_0})}(\gamma_+Z_+\to \gamma_-Z_-)$ & $\left(\frac{20}{3}d_{0,0}^0+\frac{4}{3}d_{0,0}^2\right)2\frac{f_{T_0}}{\Lambda^4} c_W^2 s_W^2 \hat{s}^2$ & $\frac{\hat{s}^2 5|f_{T_0}| c_W^2 s_W^2 }{3\pi \Lambda^4}$ \\
\hline
$ \mathcal{M}^{(f_{T_2})}(\gamma_+Z_+\to \gamma_+Z_+)$ & $6\frac{f_{T_2}}{\Lambda^4} d_{0,0}^0c_W^2 s_W^2 \hat{s}^2$ & $\frac{\hat{s}^2 3 |f_{T_2}| c_W^2 s_W^2 }{4\pi \Lambda^4}$\\
\hline
$\mathcal{M}^{(f_{T_5})}(\gamma_+Z_+\to \gamma_+Z_+)$ & $-16 \frac{f_{T_5}}{\Lambda^4} d_{0,0}^0c_W^2 s_W^2 \hat{s}^2$ & $\frac{2\hat{s}^2 |f_{T_5}| c_W^2 s_W^2 }{\pi \Lambda^4}$\\
\hline
$\mathcal{M}^{(f_{T_6})}(\gamma_+Z_+\to \gamma_+Z_+)$ & $4 \frac{f_{T_6}}{\Lambda^4} d_{0,0}^0(c_W^2-s_W^2)^2 \hat{s}^2$ & $\frac{\hat{s}^2|f_{T_6}|(c_W^2-s_W^2)^2}{2\pi \Lambda^4}$ \\
\hline
$\mathcal{M}^{(f_{T_7})}(\gamma_+Z_+\to \gamma_-Z_-)$ & $\frac{f_{T_7}}{\Lambda^4}\left(\frac{2(1-10s_W^2+10s_W^4)}{3}d_{0,0}^0 - d_{0,0}^1 + \frac{(1-2s_W^2)^2}{3}d_{0,0}^2\right) \hat{s}^2$ & $\frac{\hat{s}^2 |f_{T_7}|(10s_W^2-1-10s_W^4)}{12\pi \Lambda^4}$\\
\hline
\end{tabular}
\end{center}
\caption{\label{tab.aazzu}Same as Table~\ref{tab.wwazu} but for $\gamma Z\to \gamma Z$.}
\end{table*}
The amplitudes of $\gamma Z \to \gamma Z$ process are shown in Table~\ref{tab.aazzu}.
For those operators not listed, the amplitudes can be obtained by using
\begin{equation}
\begin{split}
&\mathcal{M}^{(O_{M_2})}(\gamma Z\to \gamma Z)=2\frac{c_W^2}{s_W^2}\frac{f_{M_2}}{f_{M_0}}\mathcal{M}^{(O_{M_0})}(\gamma Z\to \gamma Z),\\
&\mathcal{M}^{(O_{M_3})}(\gamma Z\to \gamma Z)=2\frac{c_W^2}{s_W^2}\frac{f_{M_3}}{f_{M_1}}\mathcal{M}^{(O_{M_1})}(\gamma Z\to \gamma Z),\\
&\mathcal{M}^{(O_{M_4})}(\gamma Z\to \gamma Z)=-\frac{c_W}{s_W}\frac{f_{M_4}}{f_{M_0}}\mathcal{M}^{(O_{M_0})}(\gamma Z\to \gamma Z),\\
&\mathcal{M}^{(O_{M_5})}(\gamma Z\to \gamma Z)=2\frac{c_W}{s_W}\frac{f_{M_5}}{f_{M_1}}\mathcal{M}^{(O_{M_1})}(\gamma Z\to \gamma Z),\\
&\mathcal{M}^{(O_{M_7})}(\gamma Z\to \gamma Z)=-\frac{1}{2}\frac{f_{M_7}}{f_{M_1}}\mathcal{M}^{(O_{M_1})}(\gamma Z\to \gamma Z),\\
&\mathcal{M}^{(O_{T_1})}(\gamma Z\to \gamma Z)=\frac{f_{T_1}}{f_{T_0}}\mathcal{M}^{(O_{T_0})}(\gamma Z\to \gamma Z),\\
&\mathcal{M}^{(O_{T_8})}(\gamma Z\to \gamma Z)=4\frac{f_{T_8}}{f_{T_0}}\mathcal{M}^{(O_{T_0})}(\gamma Z\to \gamma Z),\\
&\mathcal{M}^{(O_{T_9})}(\gamma Z\to \gamma Z)=4\frac{f_{T_9}}{f_{T_2}}\mathcal{M}^{(O_{T_2})}(\gamma Z\to \gamma Z).\\
\end{split}
\label{eq.a.3}
\end{equation}
The unitarity bounds are
\begin{equation}
\begin{split}
&\hat{s} ^{O_{M_0}}\leq \sqrt{\frac{128 \pi c_W^2 M_Z^2 \Lambda^4}{|f_{M_0}| e^2  v^2}},\;
 \hat{s} ^{O_{M_1}}\leq \sqrt{\frac{192 \pi c_W^2 M_Z^2 \Lambda^4}{|f_{M_1}| e^2   v^2}},\\
&\hat{s} ^{O_{M_2}}\leq \sqrt{\frac{64 \pi s_W^2 M_Z^2 \Lambda^4}{|f_{M_2}| e^2  v^2}},\;
 \hat{s} ^{O_{M_3}}\leq \sqrt{\frac{96 \pi s_W^2 M_Z^2 \Lambda^4}{|f_{M_3}| e^2   v^2}},\\
&\hat{s} ^{O_{M_4}}\leq \sqrt{\frac{128 \pi c_Ws_W M_Z^2 \Lambda^4}{|f_{M_4}| e^2  v^2}},\;
 \hat{s} ^{O_{M_5}}\leq \sqrt{\frac{96 \pi c_Ws_W M_Z^2 \Lambda^4}{|f_{M_5}| e^2   v^2}},\\
&\hat{s} ^{O_{M_7}}\leq \sqrt{\frac{384 \pi c_W^2 M_Z^2 \Lambda^4}{|f_{M_7}| e^2   v^2}},\;
 \hat{s} ^{O_{T_0}}\leq \sqrt{\frac{6 \pi \Lambda^4}{5|f_{T_0}| c_W^2 s_W^2}},\\
&\hat{s} ^{O_{T_1}}\leq \sqrt{\frac{6 \pi \Lambda^4}{5|f_{T_1}| c_W^2 s_W^2}},\;
 \hat{s} ^{O_{T_2}}\leq \sqrt{\frac{8 \pi \Lambda^4}{3 |f_{T_2}| c_W^2 s_W^2}},\\
&\hat{s} ^{O_{T_5}}\leq \sqrt{\frac{\pi \Lambda^4}{|f_{T_5}| c_W^2 s_W^2}},\;
 \hat{s} ^{O_{T_6}}\leq \sqrt{\frac{4 \pi \Lambda^4}{|f_{T_6}|(c_W^2-s_W^2)^2}},\\
&\hat{s} ^{O_{T_7}}\leq \sqrt{\frac{24 \pi \Lambda^4}{|f_{T_7}|(10s_W^2-1-10s_W^4)}},\;
 \hat{s} ^{O_{T_8}}\leq \sqrt{\frac{3 \pi \Lambda^4}{10|f_{T_8}| c_W^2 s_W^2}},\\
&\hat{s} ^{O_{T_9}}\leq \sqrt{\frac{2 \pi \Lambda^4}{3 |f_{T_9}| c_W^2 s_W^2}}.
\end{split}
\label{eq.a.4}
\end{equation}

\subsection{\label{sec:azzzvertexu}\texorpdfstring{Unitarity bounds from $ZZ\to \gamma Z$ process}{Unitarity bounds from ZZ to AZ process}}

\begin{table*}
\begin{center}
\begin{tabular}{c|c|c}
\hline
Process &  Partial wave expansion & max $|T_J|$ \\
\hline
$\mathcal{M}^{(f_{M_0})}(Z_0Z_0\to \gamma_+Z_+)$ & $-\frac{f_{M_0}}{\Lambda^4}d_{0,0}^0\frac{e^2 v^2}{2 c_W s_W M_Z^2 } \hat{s}^2$ & $\frac{\hat{s}^2 |f_{M_0}|e^2 v^2}{16\sqrt{2}\pi c_W s_W M_Z^2 \Lambda^4}$\\
\hline
$\mathcal{M}^{(f_{M_1})}(Z_0Z_0\to \gamma_+Z_+)$ & $\frac{f_{M_1}}{\Lambda^4}d_{0,0}^0\frac{e^2 v^2}{8 c_W s_W M_Z^2}\hat{s}^2$ & $\frac{\hat{s}^2 |f_{M_1}| e^2 v^2}{64\sqrt{2}\pi c_W s_W M_Z^2 \Lambda^4}$\\
\hline
$\mathcal{M}^{(f_{T_0})}(Z_+Z_+\to \gamma_-Z_-)$ & $\left(\frac{20}{3}d_{0,0}^0+\frac{4}{3}d_{0,0}^2\right)2\frac{f_{T_0}}{\Lambda^4} c_W^3 s_W \hat{s}^2$ & $\frac{\hat{s}^2 5|f_{T_0}| c_W^3 s_W }{3\sqrt{2}\pi \Lambda^4}$ \\
\hline
$ \mathcal{M}^{(f_{T_2})}(Z_+Z_+\to \gamma_+Z_+)$ & $6\frac{f_{T_2}}{\Lambda^4} d_{0,0}^0c_W^3 s_W \hat{s}^2$ & $\frac{\hat{s}^2 3 |f_{T_2}| c_W^3 s_W }{4\sqrt{2}\pi \Lambda^4}$\\
\hline
\end{tabular}
\end{center}
\caption{\label{tab.azzzu}Same as Table~\ref{tab.wwazu} but for $Z Z\to \gamma Z$.}
\end{table*}
The amplitudes of $Z Z \to \gamma Z$ process are shown in Table~\ref{tab.azzzu}.
For those operators not listed, the amplitudes can be obtained by using
\begin{equation}
\begin{split}
&\mathcal{M}^{(O_{M_2})}(Z Z\to \gamma Z)=-2\frac{f_{M_2}}{f_{M_0}}\mathcal{M}^{(O_{M_0})}(Z Z\to \gamma Z),\\
&\mathcal{M}^{(O_{M_3})}(Z Z\to \gamma Z)=-2\frac{f_{M_3}}{f_{M_1}}\mathcal{M}^{(O_{M_1})}(Z Z\to \gamma Z),\\
&\mathcal{M}^{(O_{M_4})}(Z Z\to \gamma Z)=-\frac{c_W^2-s_W^2}{2s_Wc_W}\frac{f_{M_4}}{f_{M_0}}\mathcal{M}^{(O_{M_0})}(Z Z\to \gamma Z),\\
&\mathcal{M}^{(O_{M_5})}(Z Z\to \gamma Z)=\frac{c_W^2-s_W^2}{2s_Wc_W}\frac{f_{M_5}}{f_{M_1}}\mathcal{M}^{(O_{M_1})}(Z Z\to \gamma Z),\\
&\mathcal{M}^{(O_{M_7})}(Z Z\to \gamma Z)=-\frac{1}{2}\frac{f_{M_7}}{f_{M_1}}\mathcal{M}^{(O_{M_1})}(Z Z\to \gamma Z),\\
&\mathcal{M}^{(O_{T_1})}(Z Z\to \gamma Z)=\frac{f_{T_1}}{f_{T_0}}\mathcal{M}^{(O_{T_0})}(Z Z\to \gamma Z),\\
&\mathcal{M}^{(O_{T_5})}(Z Z\to \gamma Z)=\frac{s_W^2-c_W^2}{c_W^2}\frac{f_{T_5}}{f_{T_0}}\mathcal{M}^{(O_{T_0})}(Z Z\to \gamma Z),\\
&\mathcal{M}^{(O_{T_6})}(Z Z\to \gamma Z)=\frac{s_W^2-c_W^2}{c_W^2}\frac{f_{T_6}}{f_{T_0}}\mathcal{M}^{(O_{T_0})}(Z Z\to \gamma Z),\\
&\mathcal{M}^{(O_{T_7})}(Z Z\to \gamma Z)=\frac{s_W^2-c_W^2}{c_W^2}\frac{f_{T_7}}{f_{T_2}}\mathcal{M}^{(O_{T_2})}(Z Z\to \gamma Z),\\
&\mathcal{M}^{(O_{T_8})}(Z Z\to \gamma Z)=-\frac{4s_W^2}{c_W^2}\frac{f_{T_8}}{f_{T_0}}\mathcal{M}^{(O_{T_0})}(Z Z\to \gamma Z),\\
&\mathcal{M}^{(O_{T_9})}(Z Z\to \gamma Z)=-\frac{4s_W^2}{c_W^2}\frac{f_{T_9}}{f_{T_2}}\mathcal{M}^{(O_{T_2})}(Z Z\to \gamma Z).\\
\end{split}
\label{eq.a.5}
\end{equation}
The unitarity bounds are
\begin{equation}
\begin{split}
&\hat{s} ^{O_{M_0}}\leq \sqrt{\frac{32\sqrt{2} \pi c_W s_W M_Z^2 \Lambda^4}{|f_{M_0}|e^2 v^2}},\;
 \hat{s} ^{O_{M_1}}\leq \sqrt{\frac{64\sqrt{2} \pi c_W s_W M_Z^2 \Lambda^4}{|f_{M_1}| e^2   v^2}},\\
&\hat{s} ^{O_{M_2}}\leq \sqrt{\frac{16\sqrt{2} \pi c_W s_W M_Z^2 \Lambda^4}{|f_{M_2}|e^2 v^2}},\;
 \hat{s} ^{O_{M_3}}\leq \sqrt{\frac{32\sqrt{2} \pi c_W s_W M_Z^2 \Lambda^4}{|f_{M_3}| e^2   v^2}},\\
&\hat{s} ^{O_{M_4}}\leq \sqrt{\frac{64\sqrt{2} \pi c_W^2 s_W^2 M_Z^2 \Lambda^4}{|f_{M_4}|(c_W^2-s_W^2)e^2 v^2}},\\
&\hat{s} ^{O_{M_5}}\leq \sqrt{\frac{128\sqrt{2} \pi c_W^2 s_W^2 M_Z^2 \Lambda^4}{|f_{M_5}| (c_W^2-s_W^2)e^2   v^2}},\\
&\hat{s} ^{O_{M_7}}\leq \sqrt{\frac{128\sqrt{2} \pi c_W s_W M_Z^2 \Lambda^4}{|f_{M_7}| e^2   v^2}},\;
 \hat{s} ^{O_{T_0}}\leq \sqrt{\frac{6\sqrt{2} \pi \Lambda^4}{5|f_{T_0}| c_W^3 s_W}},\\
&\hat{s} ^{O_{T_1}}\leq \sqrt{\frac{6\sqrt{2} \pi \Lambda^4}{5|f_{T_1}| c_W^3 s_W}},\;
 \hat{s} ^{O_{T_2}}\leq \sqrt{\frac{8\sqrt{2} \pi \Lambda^4}{3 |f_{T_2}| c_W^3 s_W}},\\
&\hat{s} ^{O_{T_5}}\leq \sqrt{\frac{6\sqrt{2} \pi \Lambda^4}{5|f_{T_5}| (c_W^2-s_W^2)c_W s_W}},\\
&\hat{s} ^{O_{T_6}}\leq \sqrt{\frac{6\sqrt{2} \pi \Lambda^4}{5|f_{T_6}| (c_W^2-s_W^2)c_W s_W}},\\
&\hat{s} ^{O_{T_7}}\leq \sqrt{\frac{8\sqrt{2} \pi \Lambda^4}{3 |f_{T_7}| (c_W^2-s_W^2)c_W s_W}},\;
 \hat{s} ^{O_{T_8}}\leq \sqrt{\frac{3\sqrt{2} \pi \Lambda^4}{10|f_{T_8}| c_W s_W^3}},\\
&\hat{s} ^{O_{T_9}}\leq \sqrt{\frac{2\sqrt{2} \pi \Lambda^4}{3 |f_{T_9}| c_W s_W^3}}.\\
\end{split}
\label{eq.a.6}
\end{equation}

\subsection{\label{sec:aaazvertexu}\texorpdfstring{Unitarity bounds from $\gamma \gamma \to \gamma Z$ process}{Unitarity bounds from AA to AZ process}}

\begin{table*}
\begin{center}
\begin{tabular}{c|c|c}
\hline
Process &  Partial wave expansion & max $|T_J|$ \\
\hline
$\mathcal{M}^{(f_{T_0})}(\gamma_+\gamma_+\to \gamma_-Z_-)$ & $\left(\frac{20}{3}d_{0,0}^0+\frac{4}{3}d_{0,0}^2\right)\frac{f_{T_0}}{\Lambda^4}\frac{1}{16c_W^2} \hat{s}^2$ & $\frac{\hat{s}^2 5|f_{T_0}|}{96\sqrt{2}c_W^2\pi \Lambda^4}$ \\
\hline
$ \mathcal{M}^{(f_{T_2})}(\gamma_+\gamma_+\to \gamma_+Z_+)$ & $\frac{3}{16c_W^2}\frac{f_{T_2}}{\Lambda^4} d_{0,0}^0c_W^3 s_W \hat{s}^2$ & $\frac{\hat{s}^2 3 |f_{T_2}| }{128\sqrt{2}\pi c_W^2 \Lambda^4}$\\
\hline
\end{tabular}
\end{center}
\caption{\label{tab.aaazu}Same as Table~\ref{tab.wwazu} but for $\gamma \gamma \to \gamma Z$.}
\end{table*}
The amplitudes of $\gamma \gamma \to \gamma Z$ process are shown in Table~\ref{tab.aaazu}.
For those operators not listed, the amplitudes can be obtained by using
\begin{equation}
\begin{split}
&\mathcal{M}^{(O_{T_1})}(\gamma \gamma\to \gamma Z)=\frac{f_{T_1}}{f_{T_0}}\mathcal{M}^{(O_{T_0})}(\gamma \gamma\to \gamma Z),\\
&\mathcal{M}^{(O_{T_5})}(\gamma \gamma\to \gamma Z)=\frac{c_W^2-s_W^2}{s_W^2}\frac{f_{T_5}}{f_{T_0}}\mathcal{M}^{(O_{T_0})}(\gamma \gamma\to \gamma Z),\\
&\mathcal{M}^{(O_{T_6})}(\gamma \gamma\to \gamma Z)=\frac{c_W^2-s_W^2}{s_W^2}\frac{f_{T_6}}{f_{T_0}}\mathcal{M}^{(O_{T_0})}(\gamma \gamma\to \gamma Z),\\
&\mathcal{M}^{(O_{T_7})}(\gamma \gamma\to \gamma Z)=\frac{c_W^2-s_W^2}{s_W^2}\frac{f_{T_7}}{f_{T_2}}\mathcal{M}^{(O_{T_2})}(\gamma \gamma\to \gamma Z),\\
&\mathcal{M}^{(O_{T_8})}(\gamma \gamma\to \gamma Z)=-\frac{4c_W^2}{s_W^2}\frac{f_{T_8}}{f_{T_0}}\mathcal{M}^{(O_{T_0})}(\gamma \gamma\to \gamma Z),\\
&\mathcal{M}^{(O_{T_9})}(\gamma \gamma\to \gamma Z)=-\frac{4c_W^2}{s_W^2}\frac{f_{T_9}}{f_{T_2}}\mathcal{M}^{(O_{T_2})}(\gamma \gamma\to \gamma Z).\\
\end{split}
\label{eq.a.7}
\end{equation}
The unitarity bounds are
\begin{equation}
\begin{split}
&\hat{s} ^{O_{T_0}}\leq \sqrt{\frac{192\sqrt{2} \pi c_W^2 \Lambda^4}{5|f_{T_0}|}},\;
 \hat{s} ^{O_{T_1}}\leq \sqrt{\frac{192\sqrt{2} \pi c_W^2 \Lambda^4}{5|f_{T_1}|}},\\
&\hat{s} ^{O_{T_2}}\leq \sqrt{\frac{256 \sqrt{2}\pi c_W^2 \Lambda^4}{3|f_{T_2}|}},\;
 \hat{s} ^{O_{T_5}}\leq \sqrt{\frac{192\sqrt{2} \pi c_W^2s_W^2 \Lambda^4}{5|f_{T_5}|(c_W^2-s_W^2)}},\\
&\hat{s} ^{O_{T_6}}\leq \sqrt{\frac{192\sqrt{2} \pi c_W^2s_W^2 \Lambda^4}{5|f_{T_6}|(c_W^2-s_W^2)}},\;
 \hat{s} ^{O_{T_7}}\leq \sqrt{\frac{256 \sqrt{2}\pi c_W^2s_W^2 \Lambda^4}{3|f_{T_7}|(c_W^2-s_W^2)}},\\
&\hat{s} ^{O_{T_8}}\leq \sqrt{\frac{48\sqrt{2} \pi s_W^2 \Lambda^4}{5|f_{T_8}|}},\;
 \hat{s} ^{O_{T_9}}\leq \sqrt{\frac{64 \sqrt{2}\pi s_W^2 \Lambda^4}{3|f_{T_9}|}}.\\
\end{split}
\label{eq.a.8}
\end{equation}

%\bibliography{aQGC_az}
%\bibliographystyle{h-physrev}

\end{document}